\documentclass[journal]{IEEEtran}
\usepackage{amsmath,amsfonts}
\usepackage{algpseudocode}
\usepackage{algorithm}
\usepackage{bbding}
\usepackage{array}
\usepackage[caption=false]{subfig}
\usepackage{textcomp}
\usepackage{stfloats}
\usepackage{url}
\usepackage{verbatim}
\usepackage{graphicx}
\usepackage{xcolor}
\usepackage[hidelinks]{hyperref}
\usepackage{multirow}
\usepackage{tabularx}
\usepackage{cite}

\usepackage{cleveref}
\crefname{equation}{}{}
\Crefname{equation}{Equation}{Equations}
\crefname{figure}{Fig.}{Figs.}
\crefname{table}{Table}{Tables}
\crefname{section}{Section}{Sections}
\crefname{algorithm}{Algorithm}{Algorithms}

\title{Quantum Hardware-in-the-Loop for Optimal Power Flow in Renewable-Integrated Power Systems
}
\author{Zeynab Kaseb,~\IEEEmembership{Student~Member~IEEE}, Rahul Rane,~\IEEEmembership{Student~Member~IEEE}, \\Aleksandra Leki{\'c},~\IEEEmembership{Senior~Member~IEEE}, Matthias M{\"o}ller, Amin Khodaei,~\IEEEmembership{Senior~Member~IEEE}, \\Peter Palensky,~\IEEEmembership{Senior~Member~IEEE}, and Pedro P. Vergara,~\IEEEmembership{Senior~Member~IEEE}
\thanks{
Zeynab Kaseb, Rahul Rane, Aleksandra Leki{\'c}, Peter Palensky, and Pedro P. Vergara are associated with the Intelligent Electrical Power Grids Section, Electrical Sustainable Energy, Delft University of Technology, The Netherlands. 

Matthias M{\"o}ller are associated with the Delft Institute of Applied Mathematics, Delft University of Technology, The Netherlands.

Amin Khodaei is associated with the Electrical and Computer Engineering Department at the University of Denver.

Corresponding author: Zeynab Kaseb. Email: Z.Kaseb@tudelft.nl

This study is part of the DATALESs project 482.20.602 jointly financed by the Dutch Organization for Scientific Research (NWO) and the National Natural Science Foundation of China (NSFC). The RSCAD/RTDS models and communication protocols were created as a part of the Horizon Europe project PROSECCO, under grant agreement 101160687. 
}
}

\begin{document}

\maketitle

\begin{abstract}
Quantum computing has emerged as a promising computational paradigm to address unresolved challenges in the modeling and control of modern power systems. However, most existing studies focus on offline simulations, and a practical framework for validating quantum algorithms in real-time operational environments remains lacking. This study proposes a quantum hardware-in-the-loop framework that integrates a real-time digital simulator with quantum and quantum-inspired hardware to solve combinatorial power flow and optimal power flow formulations under dynamic operating conditions. The proposed framework is validated using the IEEE 9-bus test system and a modified version with integrated solar and wind farms. The results confirm successful integration and convergence within a predefined tolerance. The study also identifies key limitations and challenges, such as limited access to quantum and digital annealers and current scalability limitations, that must be considered in future developments. Nevertheless, the results highlight the potential of quantum computing to significantly enhance the modeling and control of future power systems with high penetration of renewable energy sources.
\end{abstract}

\begin{IEEEkeywords}
Adiabatic quantum computing, combinatorial optimization, Ising machines, load flow, quantum annealing, state estimation.
\end{IEEEkeywords}

\section{Introduction}
The global transition toward sustainable energy has driven a rapid expansion of renewable energy sources (RES), particularly photovoltaic solar panels and wind turbines. However, the inherent intermittency and stochastic behavior of these resources pose significant challenges to power system operation, thus requiring advanced methodologies to ensure stability and reliability~\cite{Alam2020}. On the other hand, rising costs and environmental impacts of fossil fuels have underscored the urgency of minimizing their use and integrating RES more effectively~\cite{Asselt2024}. 

To maintain secure and efficient grid operation and assess system conditions, transmission and distribution system operators rely on power flow (PF) analysis~\cite{Kaseb2024PINN}. When economic, stability, or operational objectives must be optimized, this extends to optimal power flow (OPF). Therefore, OPF plays a crucial role in power system operation and planning by determining the most efficient system configuration through the adjustment of control variables, including active and reactive power dispatch, transformer tap positions, and reactive power compensation~\cite{Huneault1991}. Solving OPF problems enables system operators to minimize generation costs, enhance voltage stability, and ensure compliance with operational constraints, thereby improving the overall efficiency and cost-effectiveness of modern power systems~\cite{mbuli2022}.

Classical OPF solvers, including linear programming (e.g.,~\cite{Mhanna2022}), nonlinear programming (e.g.,~\cite{Momoh1999}), quadratic programming (e.g.,~\cite{Fortenbacher2019}), and Newton-based approaches (e.g.,~\cite{Silva2021}), have been extensively studied and widely adopted in academia and industry~\cite{zobaa2018}. While these solvers have proven effective for traditional power systems, their application to modern RES-integrated systems presents significant challenges. The inherent uncertainty of RES introduces complex, large-scale, and highly nonlinear and non-convex optimization problems, often leading to slow convergence and excessive computational burden~\cite{Ali2024}. As a result, advanced OPF algorithms are needed to ensure real-time applications, system stability, and optimal operation~\cite{Diab2024}.

Recent research has explored artificial intelligence (AI) algorithms as promising alternatives. These approaches employ data-driven models to enhance scalability, improve convergence speed, and enable real-time decision-making in modern power systems. For example, physics-informed neural networks (e.g.,~\cite{Misyris2019,Kaseb2024PINN}) have been employed to embed underlying equations into AI algorithms, ensuring that solutions remain physically consistent while accelerating PF analysis and OPF. Similarly, graph neural networks (e.g.,~\cite{Owerko2019}) have been applied to exploit the underlying topology of power systems, enabling efficient and scalable OPF. Reinforcement learning (e.g.,~\cite{Zeng2021}) has also garnered attention for its ability to iteratively learn optimal control policies, thus adapting to real-time power system dynamics.

Nevertheless, significant challenges remain unresolved for OPF in RES-integrated power systems. AI algorithms, while accelerating OPF, often rely on extensive training datasets and may suffer from generalization issues when encountering unseen grid conditions. In addition, their ability to guarantee feasibility and adherence to power system physics is constrained, necessitating hybrid approaches that still depend on conventional solvers. Furthermore, the non-convex, large-scale nature of OPF problems continues to pose computational challenges that classical and AI algorithms struggle to address in real-time applications (e.g.,~\cite{Narimani2018}). 

Quantum computing (QC) is a computational paradigm that exploits the principles of quantum mechanics, such as superposition and entanglement, to solve certain classes of problems more efficiently than classical computers. QC can be broadly categorized into two main paradigms: gate-based quantum computing (GQC) and adiabatic quantum computing (AQC). GQC performs computations via a sequence of quantum gates applied at discrete time steps, analogous to logical operations in classical circuits. In contrast, AQC is an analog approach that evolves a quantum system continuously toward the ground state of the problem Hamiltonian. Although AQC is polynomially equivalent to GQC, it is particularly well-suited for solving combinatorial optimization problems that can be formulated as Ising models or quadratic unconstrained binary optimization (QUBO) problems (e.g.,~\cite{kaseb2024power}).

Recently, GQC has emerged as an alternative approach for OPF, offering the potential for fast computation and improved solution accuracy. Examples are exploring quantum circuit-based methods for OPF~\cite{boddu2023quantum,hafshejani2025quantum,Gao2023}, and investigating the quantum-enhanced DC-OPF problem~\cite{amani2025quantum}. While GQC has gained significant attention due to advancements in quantum circuit design (e.g.,~\cite{feng2021quantum,feng2023noise,kaseb2024quantum}), its practical application to OPF problems remains challenging, primarily due to the dependence on large-scale, fault-tolerant hardware and the sensitivity to noise, which degrades computational accuracy and limits its effectiveness in solving large-scale OPF problems in the near to mid future (e.g.,~\cite{pareek2025limitations,liu2024quantum,magar2024dc}).

AQC exhibits better noise resilience and has demonstrated promising results on combinatorial optimization problems, which makes it a promising candidate for OPF. Examples are exploring the application of AQC to the unit commitment problem, which shares structural similarities with OPF~\cite{hartmann2025quantum}, and investigating the effectiveness of AQC for higher-order Ising models, which could be used to enhance the efficiency of OPF problems~\cite{pelofske2024short}. A comprehensive overview of the theoretical foundations and practical implementations of AQC, highlighting its advantages in solving NP-hard optimization problems, can be found in~\cite{mcgeoch2022adiabatic}.

Current AQC hardware, however, still faces limitations. For example, Wave's Advantage\texttrademark\ system (QA)\footnote{\url{https://www.dwavesys.com}} is one of the most mature realizations of AQC hardware, comprising over 5,000 qubits and 35,000 couplers, but it suffers from restricted qubit connectivity. This necessitates minor embedding to map problem graphs onto the hardware graph, a process that is NP-hard and can increase resource demands. Furthermore, the limited range of interaction coefficients and local field values can introduce precision loss during problem encoding. An alternative hardware is Fujitsu's quantum-inspired hardware (QIIO)\footnote{\url{https://www.fujitsu.com/global/services/business-services/digital-annealer}}, which emulates the annealing process using application-specific complementary metal-oxide semiconductor technology and supports massively parallel Markov Chain Monte Carlo updates. QIIO offers handling of up to 100,000 binary variables, organized in units of 8,192 fully connected bits, and operates with 64-bit precision at room temperature, thus providing a practical bridge toward scalable and noise-tolerant hardware.

Despite these promising developments, research on employing AQC for OPF remains in its early stages. Initial efforts (e.g.,~\cite{morstyn2022annealing}) have demonstrated the feasibility of AQC in power system optimization. However, challenges persist in mapping the full OPF problem onto AQC hardware. Comparative studies suggest that while AQC offers advantages for efficiently solving constrained optimization problems, further advancements in problem formulation, embedding techniques, and quantum hardware capabilities are required for practical deployment in real-time power system operations~\cite{pelofske2024short}.

While recent theoretical advances in PF analysis and OPF with AQC hardware indicate great potential~\cite{morstyn2022annealing}, their validation in real-time simulations remains essential to uncover practical limitations and implementation challenges. Hardware-in-the-loop (HIL) simulation is an effective approach for such validations, as it couples physical hardware, such as quantum or quantum-inspired processors, with real-time power system simulators. This setup provides a controlled yet realistic environment for systematically evaluating algorithmic performance, stability, and responsiveness under dynamic grid conditions, including variable renewable generation and fluctuating loads (e.g., \cite{wagle2024real,wagle2023optimal}). For example, a quantum-in-the-loop (QIL) framework that integrates GQC with real-time power system simulations is introduced in~\cite{chanda2023}, showing the potential to address complex optimization challenges in modern power systems and to enhance real-time decision-making by processing inputs from numerous distributed sensors and controllers.

This study, therefore, presents a novel quantum hardware-in-the-loop (QHIL) framework for power systems with high RES integration, where the computational capabilities of QA and QIIO are leveraged to solve combinatorial PF and OPF formulations, building on the algorithms presented in \cite{kaseb2024power}. In the proposed framework, quantum and quantum-inspired hardware are interfaced with a real-time digital simulator (RTDS\textsuperscript{\textregistered}), which accurately models and simulates real-time power system dynamics. This integration also enables the estimation of overhead time associated with QA and QIIO, which is unavoidable with the current hardware and limits potential performance in near- to mid-term applications. As a proof of concept, the framework is validated using the IEEE 9-bus test system and a modified version that incorporates solar and wind farms. The main contributions of this study are as follows:
\begin{itemize}
    \item Developing a QHIL framework that seamlessly integrates RTDS\textsuperscript{\textregistered} with AQC hardware to solve combinatorial PF and OPF formulations;
    \item Developing a Windows-based middleware that enables efficient, reliable, and real-time communication between QIIO and RTDS\textsuperscript{\textregistered}; and
    \item Systematically identifying the opportunities and challenges associated with the proposed QHIL framework, based on proof-of-concept simulations of a small-scale power system achieving predefined accuracy thresholds for PF and OPF.
\end{itemize}

\section{Mathematical Formulation}\label{sec:math_formulation}

\subsection{Power Flow (PF) Analysis}
The objective is to compute the bus voltages and power injections that satisfy the power balance equations, as:
\begin{subequations} \label{eq:pf-cons}
    \begin{align}
        P_i & = P_i^G - P_i^D, \quad \forall i \in \{1,\dots,N\}, \label{eq:active-power-balance}\\
        Q_i & = Q_i^G - Q_i^D, \quad \forall i \in \{1,\dots,N\}, \label{eq:reactive-power-balance}
    \end{align}
\end{subequations}
where $P_i$ and $Q_i$ are net active and reactive power injections; $P_i^G$ and $Q_i^G$ are total generated active and reactive power; $P_i^D$ and $Q_i^D$ total demand active and reactive power.

Equation \cref{eq:pf-cons} is typically solved iteratively using numerical methods, such as the NR or Gauss-Seidel. It can also be solved with quantum or quantum-inspired (digital) annealers, which amounts to minimizing the sum of all the terms squared as:
\begin{equation} \label{eq:H_obj}
    H_{\text{obj}}(\mathbf{x}) = \sum_{i=1}^{N} (P_i - P_i^G + P_i^D)^2 + (Q_i - Q_i^G + Q_i^D)^2 ,
\end{equation}
where $\mathbf{x} \in \{0,1\}$ is a vector of binary variables. To convert the problem into a compatible form with quantum/digital annealers, $P_i$ and $Q_i$ are presented in rectangular coordinates and expanded into:
\begin{subequations} \label{eq:pq-sum_expanded}
    \begin{align}
        P_i & = \sum_{k=1}^{N} \mu_i G_{ik} \mu_k + \omega_i G_{ik} \omega_k + \omega_i B_{ik} \mu_k - \mu_i B_{ik} \omega_k , \label{eq:p-sum_expanded}\\ 
        Q_i & = \sum_{k=1}^{N} \omega_i G_{ik} \mu_k - \mu_i G_{ik} \omega_k - \mu_i B_{ik} \mu_k - \omega_i B_{ik} \omega_k, \label{eq:q-sum_expanded}
    \end{align}
\end{subequations}
where $\mu_i$ and $\omega_i$ are the real and imaginary parts of the complex voltage at bus $i$; $G_{ik}$ and $B_{ik}$ are the conductance and susceptance between buses $i$ and $k$.

\subsection{Optimal Power Flow (OPF)}
The objective is to compute the optimal generator outputs $\{P_i^G, Q_i^G\}$ that minimize generation costs while satisfying all system constraints, i.e., the power balance, generation, and operational limits, formulated as:
\begin{equation} \label{eq:opf-obj}
    \min \sum_{k \in \mathbb{G}} f_k(P_k^G),
\end{equation}
subject to:
\begin{subequations} \label{eq:opf-cons}
    \begin{align}
        P_i & = P_i^G - P_i^D, \label{eq:opf-active-power-balance}\\
        Q_i & = Q_i^G - Q_i^D, \label{eq:opf-reactive-power-balance}\\
        \underline{P_i^G} & \leq P_i^G \leq \overline{P_i^G}, \label{eq:active-power-limits}\\
        \underline{Q_i^G} & \leq Q_i^G \leq \overline{Q_i^G}, \label{eq:reactive-power-limits}\\
        \underline{V_i} & \leq V_i \leq \overline{V_i}, \label{eq:voltage-limits}\\
        \underline{\delta_i} & \leq \delta_i \leq \overline{\delta_i}, \label{eq:angle-limits}
    \end{align}
\end{subequations}
where $f_k(\cdot)$ is the fuel cost function; $\mathbb{G}$ is a subset of buses connected to generators \{1, 2, ..., $N_G$\}; $N_G$ is the total number of generators; $\underline{P_i^G}$ and $\overline{P_i^G}$ are the minimum and maximum limits for the generated active power at bus $i$; $\underline{Q_i^G}$ and $\overline{Q_i^G}$ are the minimum and maximum limits for the generated reactive power at bus $i$; ${V_i}$ is the voltage magnitude at bus $i$; $\underline{V_i}$ and $\overline{V_i}$ are the minimum and maximum limits for the voltage magnitude at bus $i$; ${\delta_i}$ is the voltage phase angle at bus $i$; $\underline{\delta_i}$ and $\overline{\delta_i}$ are the minimum and maximum limits for the voltage phase angle at bus $i$.

We restructure the OPF formulation into a problem Hamiltonian, including the objective term, $H_{\text{obj}}$, as defined in \cref{eq:H_obj}, along with the inequality constraint terms, $H_{\text{const}}$, and the quality constraint term, $H_{\text{cost}}$, as defined in \cref{eq:H_const,eq:H_cost}, respectively.

Inequality constraints \cref{eq:active-power-limits,eq:reactive-power-limits,eq:voltage-limits,eq:angle-limits} are incorporated into the problem Hamiltonian as penalty terms. A typical penalty term for an inequality constraint $g(x) \leq 0$ is $\lambda \max(0, g(x))^2$, where $\lambda$ is a penalty parameter, resulting in: 
\begin{equation} \label{eq:H_const}
    \begin{aligned}
        H_{\text{const}}(\mathbf{x}) = \sum_{i \in \mathbb{G}} \Big[ & \lambda_{0} \max(0, P_i^G - \overline{P_i^G})^2\\
        + & \lambda_{1} \max(0, \underline{P_i^G} - P_i^G)^2\\[4px]
        + & \lambda_{2} \max(0, Q_i^G - \overline{Q_i^G})^2\\[4px]
        + & \lambda_{3} \max(0, \underline{P_i^G} - Q_i^G)^2 \Big] \\
        + \sum_{i=1}^{N} \Big[ & \lambda_{4} \max(0, V_i - \overline{V_i})^2\\
        + & \lambda_{5} \max(0, \underline{V_i} - V_i)^2\\[4px]
        + & \lambda_{6} \max(0, \delta_i - \overline{\delta_i})^2\\[4px]
        + & \lambda_{7} \max(0, \underline{\delta_i} - \delta_i)^2 \Big] .
    \end{aligned}
\end{equation}

Finally, the objective function \eqref{eq:opf-obj} is incorporated into the problem Hamiltonian as a penalty term. A typical penalty term for an equality constraint $h(x) = 0$ is $\lambda h(x)^2$, where $\lambda$ is a penalty parameter, resulting in:
\begin{equation} \label{eq:H_cost}
    H_{\text{cost}}(\mathbf{x}) = \lambda_{8} \sum_{k \in \mathbb{G}} f_k(P_k^G)^2.
\end{equation}

Combining the objective and constraints gives:
\begin{equation}
    \label{eq:final-hamiltonian}
    H(\mathbf{x}) = H_{\text{obj}}(\mathbf{x}) + H_{\text{const}}(\mathbf{x}) + H_{\text{cost}}(\mathbf{x}).
\end{equation}

While \cref{eq:H_const,eq:H_cost} allow for encoding constraints within the problem Hamiltonian, current AQC hardware does not natively support $\max(\cdot)$ operations, nor can they directly enforce inequality conditions. Therefore, we introduce slack binary variables to transform these constraints into an equivalent binary form compatible with existing implementations. Although this transformation increases the problem size, it offers a practical and currently necessary means to represent operational limits within the combinatorial formulation. The trade-off between model fidelity and hardware feasibility is therefore an inherent aspect of current implementations. 

\subsection{QUBO Representation}
To further transform \cref{eq:H_obj,eq:final-hamiltonian} into a form compatible with AQC hardware, the continuous variables \(\mu_i\) and \(\omega_i\) in \cref{eq:pq-sum_expanded} must be discretized. A straightforward discretization is:
\begin{subequations} \label{eq:muomega-increment}
    \begin{align}
        \mu_i & = \mu_i^0 + \Delta \mu_i (x_{i,0}^\mu - x_{i,1}^\mu), \label{eq:mu-increment}\\
        \omega_i & = \omega_i^0 + \Delta \omega_i(x_{i,0}^\omega - x_{i,1}^\omega), \label{eq:omega-increment}
    \end{align}
\end{subequations}
where $x_{i,\{0,1\}}^{\{\mu,\omega\}}\in\{0,1\}$ are binary decision variables whose value decides whether the base values $\mu_i^0$ and $\omega_i^0$ are increased
($x_{i,0}^{[\circ]} = 1 \,\land\, x_{i,1}^{[\circ]} = 0$), decreased ($x_{i,0}^{[\circ]} = 0 \,\land\, x_{i,1}^{[\circ]} = 1$), or kept at their current value ($x_{i,0}^{[\circ]} = 0 \,\land\, x_{i,1}^{[\circ]} = 0$ or $x_{i,0}^{[\circ]} = 1 \,\land\, x_{i,1}^{[\circ]} = 1$). It should be noted that $[\circ]$ is a placeholder notation for $\mu$ and $\omega$, respectively. That is, $\mu_i^0$ can increase ($x_{i,0}^\mu = 1 \,\land\, x_{i,1}^\mu = 0$) while $\omega_i^0$ decreases ($x_{i,0}^\omega = 0 \,\land\, x_{i,1}^\omega = 1$).

Working out all the terms at \cref{eq:H_obj,eq:final-hamiltonian} yields a fourth-order polynomial for the binary variables. Detailed information on the expanded formulations is available in \cite{kaseb2024power}. The solvers are QA and QIIO. While QIIO can effectively handle higher-order terms, QA can handle up to quadratic terms. Therefore, higher-order terms are reduced by introducing auxiliary variables of the form $z_{ij}=x_i x_j$ and replacing triplet interactions by:
\begin{equation}
    x_i x_j x_k = \min_{z_{ij}} \left\{z_{ij}x_k + \lambda P(x_i, x_j; z_{ij})\right\},\quad \lambda > 0,
\end{equation}
with:
\begin{equation}
    P(x_i, x_j; z_{ij}) = x_i x_j - 2(x_i+x_j)z_{ij} + 3z_{ij},
\end{equation}

Similarly, expressions with four binary variables can be replaced by:
\begin{equation}
    x_i x_j x_k x_l = \min_{z{ij}, z_{kl}} \left\{\lambda P(x_i, x_j; z_{ij}) + \lambda P(x_k, x_l; z_{kl})\right\}.
\end{equation}

The discretized formulation can then be expressed as a QUBO formulation, which is minimized using quantum or quantum-inspired hardware. A QUBO formulation is defined by a symmetric, real-valued matrix $\boldsymbol{Q} \in \mathbb{R}^{n \times n}$, where $n$ is the total number of decision variables, including slack and auxiliary variables. The corresponding binary optimization problem is formulated as:
\begin{equation}
    \min_{\mathbf{x} \in \{0,1\}^n} f_{\boldsymbol{Q}}(\mathbf{x}),
\end{equation}
with the quadratic objective function given by:
\begin{equation}
    f_{\boldsymbol{Q}}(\mathbf{x}) = \mathbf{x}^\top \boldsymbol{Q} \mathbf{x} = \sum_{i=1}^{n} \sum_{j=1}^{n} Q_{ij} x_i x_j. \label{eq:qubo}
\end{equation}

The Python package PyQUBO is used to develop the QUBO based on \cref{eq:H_obj} for PF and \cref{eq:final-hamiltonian} for OPF when the solver is QA to reduce higher-order terms.

\section{Algorithm Implementation} \label{sec:implementation}

A detailed description of the designed algorithm for PF analysis can be found in \cite{kaseb2024power}. \cref{alg:aqopf} is designed to solve the combinatorial OPF formulation \cref{eq:final-hamiltonian} using quantum and digital annealers. In lines 1-3, $\mathbf{P}^{D}$, $\mathbf{Q}^{D}$, and $\mathbf{Y}=G_{ik}+\textrm{j}B_{ik}$ are defined based on the given operating scenario. In lines 4–7, $\mathbf{\mu}^0$, $\mathbf{\omega}^0$, $\mathbf{\Delta{\mu}}$, and $\mathbf{\Delta{\omega}}$ are initialized by the user. The values of $\mathbf{P}$ and $\mathbf{Q}$ are then calculated in lines 8 and 9 using $\mathbf{\mu}^0$ and $\mathbf{\omega}^0$, where $P_1$ and $Q_1$ associated with the \emph{slack} bus are excluded. In line 10, the problem Hamiltonian $H(\cdot)$ is evaluated with $\mathbf{x}=\{0\}^{4n}$ to propose a candidate solution with the initial values for $\mathbf{\mu}^0$, $\mathbf{\omega}^0$, $\mathbf{\Delta{\mu}}$, and $\mathbf{\Delta{\omega}}$. In line 11, the residual threshold $\epsilon$ $\frac{{\text{MW}^2 + \text{MVAR}^2}}{2}$ is defined by the user. In line 12, the iteration counter $it$ is set to zero.

\begin{algorithm}[t]
\caption{Adiabatic quantum algorithm to solve the combinatorial OPF formulation.}\label{alg:aqopf}
\begin{algorithmic}[1]
    \State Initialize $\mathbf{P}^{D} = [P^{D}_1, P^{D}_2, \dots, P^{D}_{N - N_G - 1}]$
    \State Initialize $\mathbf{Q}^{D} = [Q^{D}_1, Q^{D}_2, \dots, Q^{D}_{N - N_G - 1}]$
    \State Initialize $\mathbf{Y}= \{(G_{ik}+\textrm{j}B_{ik}): i,k=1,2,\dots,N\}$
    \State $\mathbf{\Delta{\mu}} \gets 1 \times 10^{-2}$
    \State $\mathbf{\Delta{\omega}} \gets 1 \times 10^{-3}$ 
    \State $\mathbf{\mu}^0 = [\mu_{1}^0, \mu_{2}^0, \dots, \mu_{N}^0] \gets 1$
    \State $\mathbf{\omega}^0 = [\omega_{1}^0, \omega_{2}^0, \dots, \omega_{N}^0] \gets 0$
    \State Calculate $\mathbf{P}$ for \emph{PQ} buses using \cref{eq:p-sum_expanded}
    \State Calculate $\mathbf{Q}$ for \emph{PQ} buses using \cref{eq:q-sum_expanded}
    \State Calculate residual using $H(\mathbf{x})$ in \cref{eq:final-hamiltonian}
    \State $\epsilon \gets 1 \times 10^{-2}$
    \State $it \gets 0$
    \While{$H_{obj}(\mathbf{x}) > \epsilon$ and $it<it_{\text{max}}$}
        \State Solve QUBO and Update $\mathbf{x}$
        \State Update $\mathbf{\mu}^0$ and $\mathbf{\omega}^0$ using \cref{eq:muomega-increment}
        \State Calculate $\mathbf{P}$ for \emph{PQ} buses using \cref{eq:p-sum_expanded}
        \State Calculate $\mathbf{Q}$ for \emph{PQ} buses using \cref{eq:q-sum_expanded}
        \State Calculate residual using $H(\mathbf{x})$ in \cref{eq:final-hamiltonian}
        \State Update $\mathbf{\mu}^0$ and $\mathbf{\omega}^0$ 
        \State Update $\mathbf{\Delta{\mu}}$, $\mathbf{\Delta{\omega}}$
        \State $it \gets it + 1$
    \EndWhile
\end{algorithmic}
\end{algorithm}

The iterative loop starts in line 13. In the first iteration, the initial candidate solution $\mathbf{x}=\{0\}^{4n}$ is used to update $\mathbf{\mu}^0$ and $\mathbf{\omega}^0$ in line 15, followed by calculating $\mathbf{P}$ and $\mathbf{Q}$ in lines 16 and 17. From the second iteration on, after solving the QUBO problem in line 14 using QA or QIIO, the resulting bitstring $\mathbf{x}=[x_{0,0}^\mu,\dots,x_{n,1}^\omega] \in \{0,1\}^{4n}$ is obtained and used to update $\mathbf{\mu}^0$ and $\mathbf{\omega}^0$ in line 15 according to \cref{eq:muomega-increment}. Note that the auxiliary variables are neglected. For all solvers, the number of readouts is 2,000. When the calculated residual using $H(\mathbf{x})$ in line 18 falls below $\epsilon$, $\mathbf{\mu}^0$, $\mathbf{\omega}^0$ are considered as the final solution. If not, they are updated in line 19.

In line 20, the step sizes are adaptively updated to enhance convergence. Specifically, $\mathbf{\Delta{\mu}}$ and $\mathbf{\Delta{\omega}}$ are defined as functions of $it$, such that larger step sizes are applied during the initial iterations to accelerate convergence, while smaller step sizes are used as the algorithm approaches the solution. Furthermore, $\mathbf{\Delta{\mu}}$ and $\mathbf{\Delta{\omega}}$ are updated per bus considering the history of changes in $\mathbf{\mu}^0$ and $\mathbf{\omega}^0$ from the previous and second previous iterations. This approach enables independent adjustment of $\mu_i$ and $\omega_i$ in different directions. The loop continues until convergence or the maximum iteration count $it_{\text{max}}$ is reached.

\section{Quantum Hardware-in-the-Loop Framework} \label{sec:qhil_framework}

\subsection{Computational and Information Flow}
\begin{figure}[t!]
    \centering
\includegraphics[width=0.9\columnwidth]{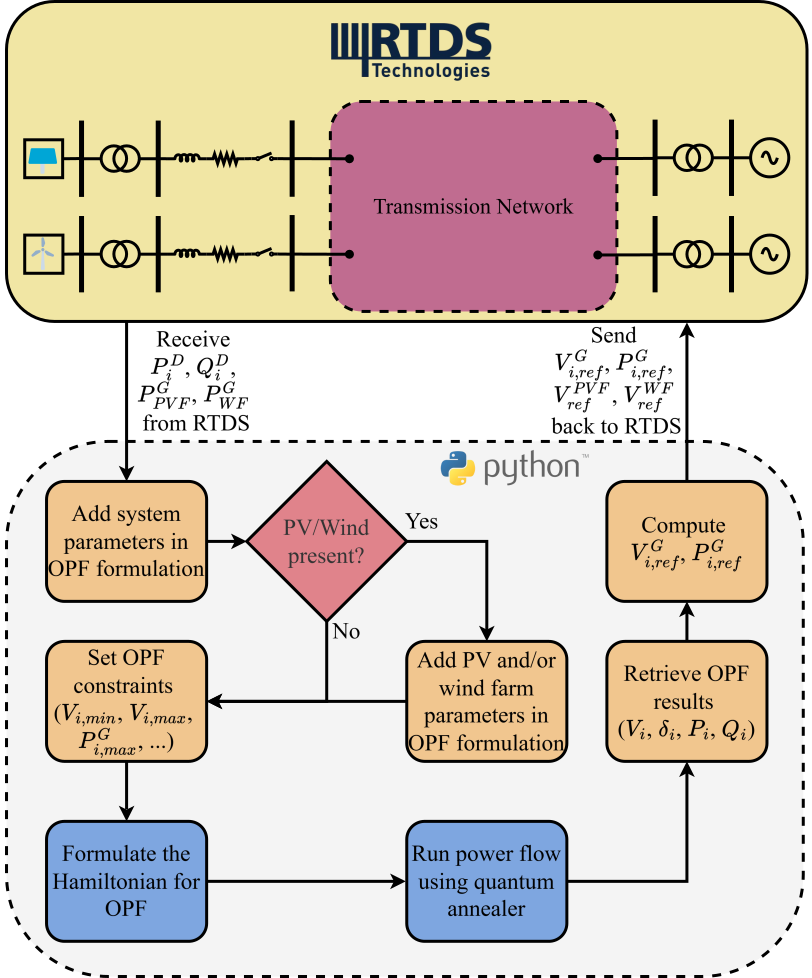}
    \caption{Computational and information flow within the QHIL framework for OPF calculations. The OPF formulation receives system parameters, real-time data, $P_i^D$, $Q_i^D$, $P_i^{PVF}$, $P_i^{WF}$, and operational constraints. The OPF is solved using a problem Hamiltonian on quantum/digital annealers. The results are compared to the NR solver in Pandapower. The computed generator set points $V_{i,ref}^G$ and $P_{i,ref}^G$ are then used in the real-time power system simulation.}
    \label{fig:QHIL_Loop}
\end{figure}

\cref{fig:QHIL_Loop} depicts the computational and information flow within the QHIL framework for OPF. Different colors and line styles are used to distinguish components and improve visual clarity. In particular, the yellow box represents the RTDS\textsuperscript{\textregistered} hardware, while the dashed gray box corresponds to the Python environment. Solid box outlines denote the overall system architecture, whereas dashed boxes indicate internal processes, with arrows passing through them to show step-by-step operations and data flows within the system. The problem formulation is initially given essential power system parameters, such as the number of buses $N$, system connectivity, and transmission line resistance, inductance, and capacitance. In addition, real-time data are obtained from the simulator, including demand power values $P_i^D$ and $Q_i^D$, as well as RES power generation outputs $P_i^{PVF}$ and $P_i^{WF}$. If solar and/or wind farms are integrated into the power system, the corresponding parameters are added as well. For OPF, relevant constraints, including bus voltage limits $\underline{V_i}$ and $\overline{V_i}$, as well as generation limits $\underline{P_i^G}$, $\overline{P_i^G}$, $\underline{Q_i^G}$, and $\overline{Q_i^G}$, are provided to ensure that the power system functions within acceptable operating limits.

The OPF problem is then solved using the proposed algorithm in \cref{sec:implementation}, executed on QA and QIIO. The results are subsequently collected from the annealers and can be used for modeling and control purposes, e.g., to calculate the required generator voltage $V_{i,ref}^G$ and active power $P_{i,ref}^G$ set points based on the approach used for modeling the generator's excitor and governor in the real-time power system simulation. 

The required generator voltage $V_{i,ref}^G$ set point is given by:
\begin{subequations} \label{eq:vref_gen}
    \begin{align}
        V_{i,ref}^G & = \frac{E^G_{f,i,pu}}{K_a} + V^G_{i,pu} \\
        \Bar{E}^G_{f,i,pu} & = \Bar{V}^G_{i,pu} + jx_d \Bar{I}^G_{i,pu} \\
        \Bar{I}^G_{i,pu} & = \frac{S_{i,pu}^*}{\Bar{V}^{G*}_{i,pu}} \\
        S_{i,pu} & = \frac{P_i^G + jQ_i^G}{S_{base}}
    \end{align}
\end{subequations}
where $K_a$ is the excitor gain; $x_d$ the $d$-axis reactance of the generator; $\Bar{E}^G_{f,i,pu}$ the field voltage of the generator in per-unit; $\Bar{I}^G_{i,pu}$ the current flowing through the generator in per-unit; $\Bar{V}^G_{i,pu}$ the generator terminal voltage; $S_{i,pu}$ the apparent power supplied by the generator in per-unit; $S_{base}$ the base MVA value. 

Similarly, the generator's active power $P_{i,ref}^G$ set point can be computed as:
\begin{equation} \label{eq:pref_gen}
    P_{i,ref}^G = R\frac{P_i^G}{S_{base}}
\end{equation}
where $R$ is the governor droop coefficient. 

Finally, the estimated generator and RES converter set points $V_{i,ref}^G$, $P_{i,ref}^G$, $V_{i,ref}^{PVF}$, and $V_{i,ref}^{WF}$ are returned to the simulator as inputs for system operation.

\subsection{Hardware Configuration}
\begin{figure}[t!]
    \centering
\includegraphics[width=0.9\columnwidth]{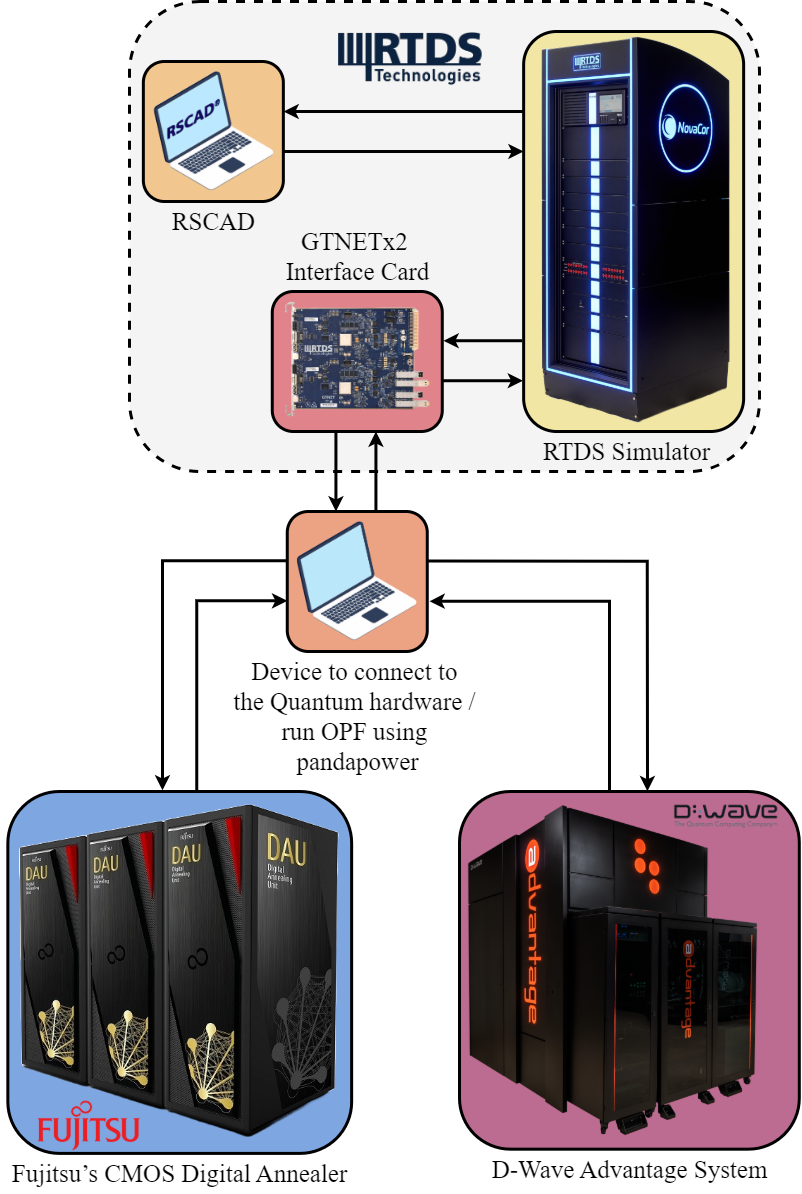}
    \caption{Hardware configuration of the proposed QHIL framework for real-time power system simulations using RSCAD/RTDS\textsuperscript{\textregistered}. A GTNETx2 card with SKT protocol enables TCP communication to exchange system parameters and RES generation data with the OPF algorithm running on the quantum hardware or local device, which then returns generator and converter set points to the simulator in real time.}
    \label{fig:QHIL}
\end{figure}

\cref{fig:QHIL} shows the hardware configuration of the QHIL framework, where the dashed gray box highlights all equipment modeled in RTDS\textsuperscript{\textregistered} and the associated connections, showing the internal operations of the real-time simulation framework. Power system simulations are carried out on the RTDS\textsuperscript{\textregistered} NovaCor racks, and responses can be monitored using the RSCAD interface\footnote{\url{https://www.rtds.com/technology/graphical-user-interface}}. To provide real-time connectivity with the simulator, an external communication GTNETx2 card is used. When configured with the card's SKT protocol, it allows the simulator and external devices to communicate via the Transmission Control Protocol (TCP). This TCP communication further allows the simulator to export system parameters, such as power demand values $P_i^D$ and $Q_i^D$, as well as RES power generation $P_i^{PVF}$ and $P_i^{WF}$ for OPF. Here, $P_i^{PVF}$ and $P_i^{WF}$ are the active power generated by the solar and wind farms, respectively. Subsequently, the computed generator and RES converter set points $V_{i,ref}^G$, $P_{i,ref}^G$, $V_{i,ref}^{PVF}$, and $V_{i,ref}^{WF}$ are sent back to the simulator in real-time. 

RTDS\textsuperscript{\textregistered} can be directly interfaced with QA. However, a direct connection to QIIO is currently not feasible, primarily due to the lack of proximity between the hardware and the simulation environment, and secondly, because vendor-specific VPN network access is required. Therefore, a Windows-based middleware is developed to facilitate communication between QIIO and RTDS\textsuperscript{\textregistered}.

The inputs required for the OPF problem are sent to the proposed algorithms in \cref{sec:implementation} via Python scripts. The Python script is executed in a Windows Subsystem for Linux (WSL) environment, which runs a virtual Ubuntu instance, where the required Python packages (e.g., D-Wave SDK, PyQUBO, DADK) are installed. Azure VPN Client is used to establish a connection with QIIO. However, this step is not necessary for QA. After obtaining solutions for OPF, the corresponding generator set points are then transmitted back to the simulation loop.

\section{Results} \label{sec:results}
The proposed QHIL framework is validated by performing PF analysis and OPF for the IEEE 9-bus test system and a modified version that incorporates solar and wind farms, resulting in a 13-bus test system. Power system simulations are performed on the RTDS\textsuperscript{\textregistered} NovaCor rack, and AQC experiments are performed on QA and QIIO. The compiler used for the former is PyQUBO, and for the latter is DADK. The tolerance threshold of $\epsilon = 1 \times 10^{-2}$ is chosen for the error bounds of PF and OPF due to limited hardware access and the computational expenses associated with QA and QIIO. As observed, the residuals obtained at termination remain close to the prescribed threshold, indicating that the iterative process is approaching convergence and would reach tighter tolerances if additional iterations were permitted. Therefore, the imposed iteration cap reflects high computational cost and limited access to available quantum(-inspired) computing resources, rather than the algorithm's theoretical convergence behavior or scalability.

\subsection{Computational Performance Assessment}
The computational setup for solving PF and OPF combinatorial formulations using QA and QIIO is summarized in \cref{tab:solvers}. Key computational performance metrics include the number of variables in the QUBO formulation, compilation time, iteration time, and the total number of iterations required for convergence. Overall, QIIO demonstrates significantly faster compilation and iteration times compared to QA. For example, in the PF problem for the IEEE 9-bus test system, the compilation time drops from 4.35 seconds with QA to 0.27 seconds with QIIO, while the iteration time reduces from 238.48 seconds with QA to 0.93 seconds with QIIO. The OPF problem for the 13-bus test system exhibits higher computational complexity, involving 635 variables and a longer compilation time of 41.25 seconds.

The computational times reported in \cref{tab:solvers} include the computation overhead associated with QA and QIIO, which is unavoidable with current hardware. The processing times for QA and QIIO are also compared with that of the NR solver. For the 9-bus test system, NR achieves an average processing time of 0.049 seconds for PF and 0.397 seconds for OPF, measured over 10,000 runs. For 2,000 readouts/samples per iteration, the average anneal time per iteration for PF is 0.04 seconds using QA and 0.671 seconds using QIIO. The corresponding time for OPF on the 9-bus test system using QIIO is 0.981 seconds. Note that with NR, the OPF processing time is roughly 8 times that of PF, while with QIIO, this ratio is less than 2, which suggests improved scalability potential as hardware continues to advance. It should be emphasized, however, that the processing times reported for QA and QIIO are subject to variability due to several factors, such as hardware noise, specialized interfaces, and VPN connections. Consequently, accurately isolating the pure processing time remains challenging at present.

\begin{table*}[t!]
    \centering
    \caption{Performance comparison of QA and QIIO for the PF and OPF problems in terms of variable count, compilation time, and iteration time across various test systems.}
    \label{tab:solvers}
    \begin{tabular}{cccccccc}
        \hline
        Test & Problem & Annealer & Compiler & Var. & Compile & It. & It. \\
        System & & & & No. & Time (s) & Time (s) & No. \\
        \hline
         9-bus  & PF  & QA   & PyQUBO & 666 & 4.35  & 238.48 & 152 \\
         9-bus  & PF  & QIIO & DADK   & 249 & 0.27  & 0.93   & 119 \\
         9-bus  & OPF & QIIO & DADK   & 331 & 0.42  & 1.81   & 200 \\
         13-bus & OPF & QIIO & DADK   & 635 & 41.25 & 42.37  & 200 \\
        \hline
    \end{tabular}
\end{table*}

\subsection{Error Assessment}
\cref{tab:solvers_results} shows the differences in $V_i$, $\delta_i$, $P_i$, and $Q_i$ for PF and OPF using QA and QIIO relative to the results obtained by NR. These differences are quantified as the mean and standard deviation of the absolute discrepancies between the reference NR solution and the QA or QIIO solutions across all buses. Accordingly, the final deviations for $V_i$ and $\delta_i$ remain within the predefined residual threshold ($10^{-2}$), which indicates the feasibility and reliability of the proposed algorithms.

\begin{table*}[t!]
    \centering
    \caption{Deviation analysis of values for $V_i$, $\delta_i$, $P_i$, and $Q_i$ obtained by QA and QIIO for the PF and OPF problems.}
    \label{tab:solvers_results}
    \begin{tabular}{ccccccccccc}
        \hline
        Test & Problem & Annealer & $\overline{|\Delta V_i|}$ & $\sigma_{|\Delta V_i|}$ & $\overline{|\Delta \delta_i|}$ & $\sigma_{|\Delta \delta_i|}$ & $\overline{|\Delta P_i|}$ & $\sigma_{|\Delta P_i|}$ & $\overline{|\Delta Q_i|}$ & $\sigma_{|\Delta Q_i|}$ \\
        System & & & (p.u.) & (p.u.) & (deg.) & (deg.) & (MW) & (MW) & (MVAR) & (MVAR) \\
        \hline
         9-bus  & PF  & QA & $2.53 \times 10^{-4}$ & $2.41 \times 10^{-4}$ & $2.13 \times 10^{-3}$ & $1.37 \times 10^{-3}$ & 0.03 & 0.04 & 0.18 & 0.29 \\
         9-bus  & PF  & QIIO  & $1.29 \times 10^{-3}$ & $1.64 \times 10^{-3}$ & $1.88 \times 10^{-2}$ & $2.15 \times 10^{-2}$ & 0.47 & 0.76 & 0.86 & 1.85 \\
         9-bus  & OPF & QIIO & $1.24 \times 10^{-3}$ & $1.75 \times 10^{-3}$ & $1.37 \times 10^{-2}$ & $9.95 \times 10^{-3}$ & 0.13 & 0.23 & 1.82 & 3.25 \\
         13-bus & OPF & QIIO & $1.57 \times 10^{-3}$ & $9.83 \times 10^{-4}$ & $1.52 \times 10^{-2}$ & $2.20 \times 10^{-2}$ & 0.06 & 0.16 & 1.13 & 2.84 \\
        \hline
    \end{tabular}
\end{table*}

\begin{figure*}[htbp] 
    \centering 
    \subfloat[\label{fig:Vi_qc_pp_9bus_pf}]{
        \begin{minipage}{0.24\textwidth}
            \centering
            \includegraphics[width=\columnwidth]{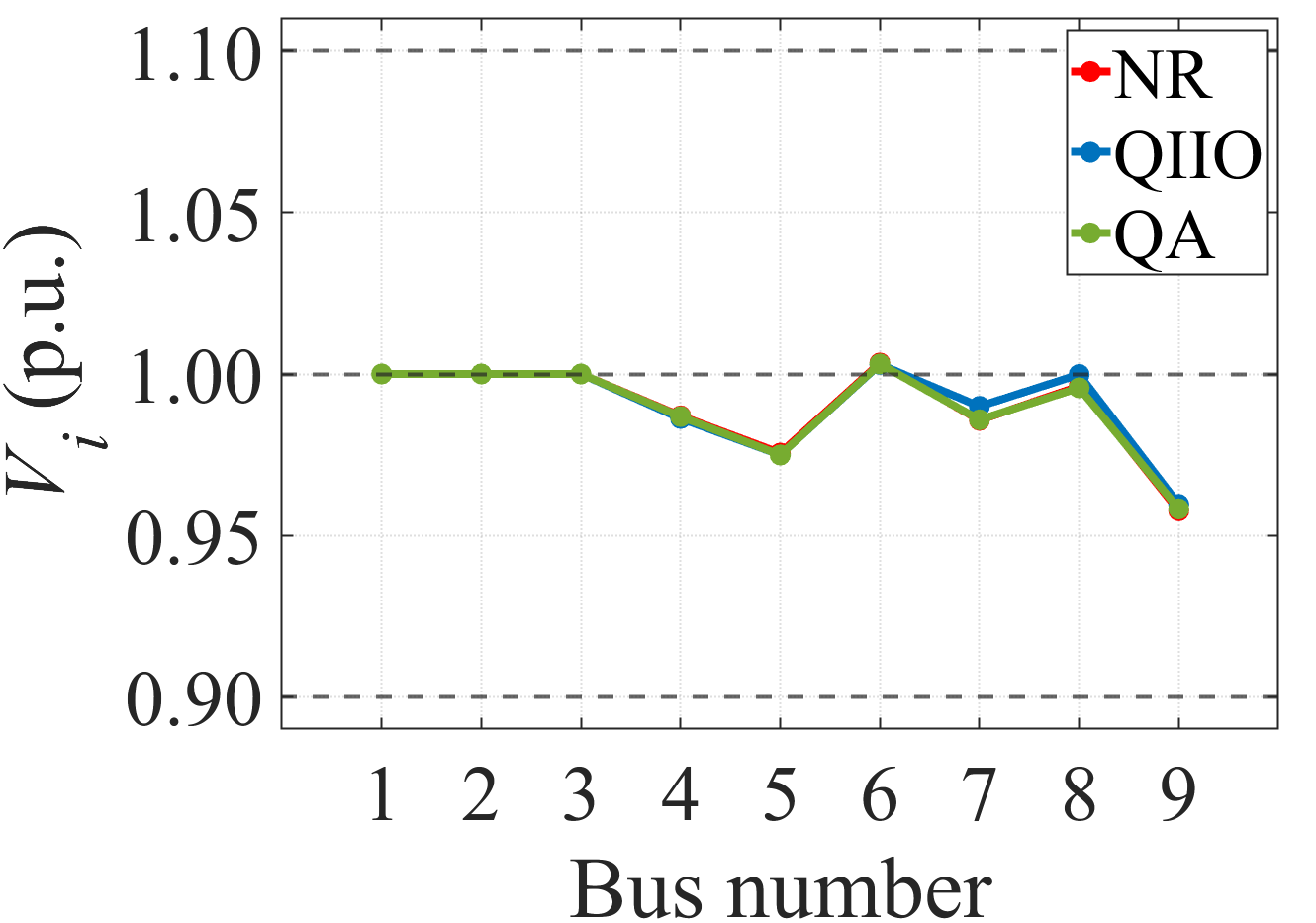} \\
            \includegraphics[width=\columnwidth]{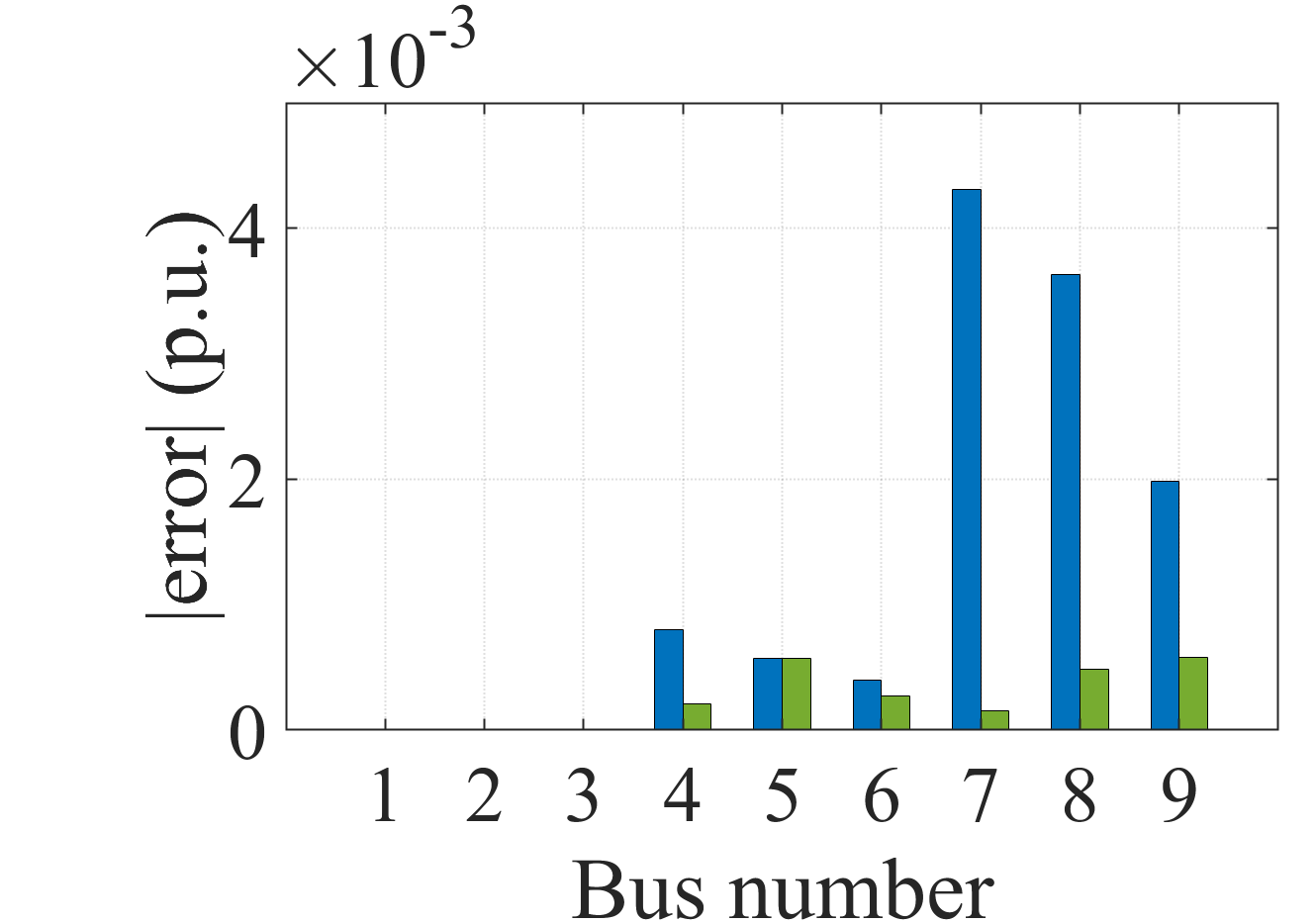}
        \end{minipage}
    }%
    \subfloat[\label{fig:deltai_qc_pp_9bus_pf}]{
        \begin{minipage}{0.24\textwidth}
            \centering
            \includegraphics[width=\columnwidth]{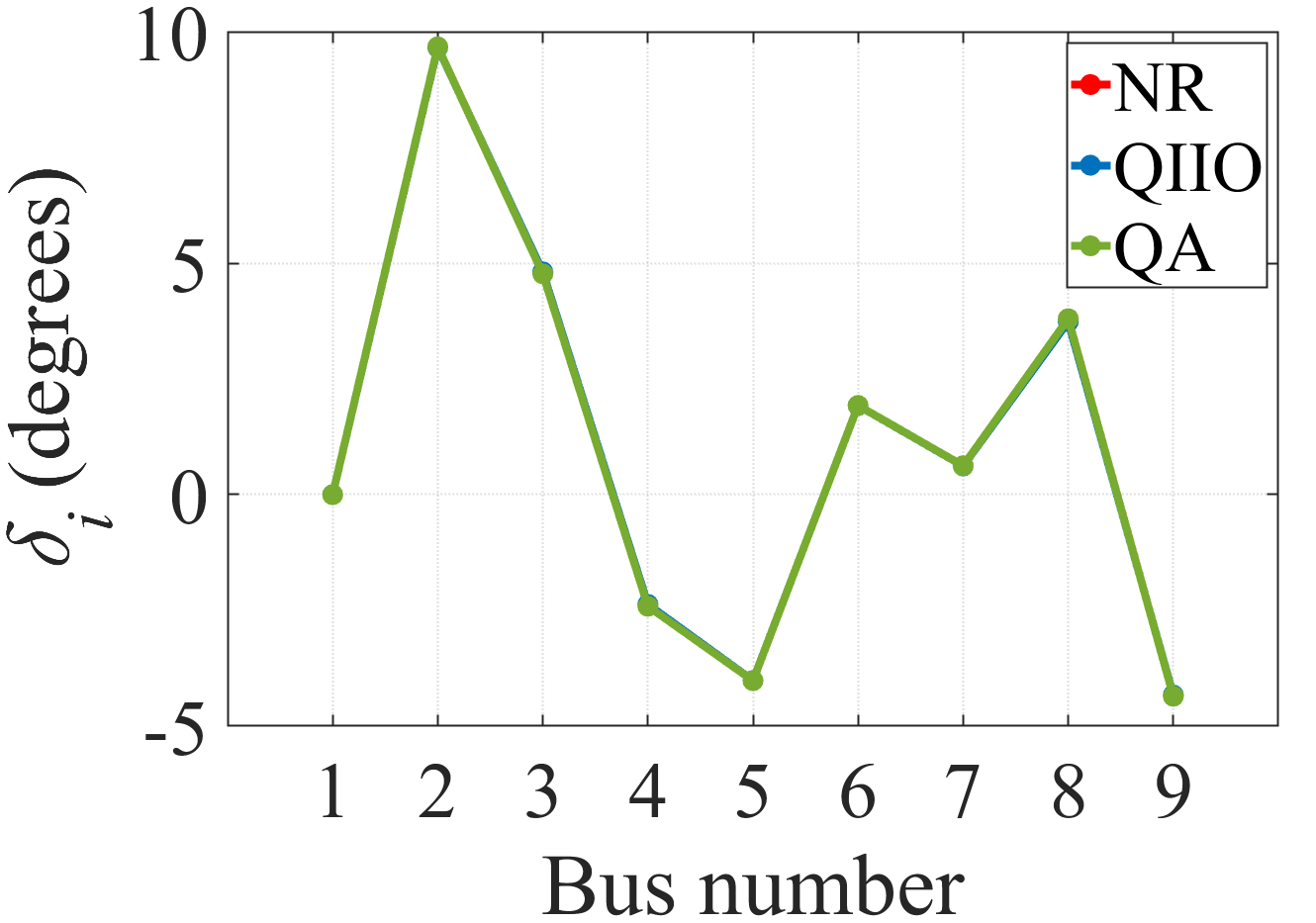} \\
            \includegraphics[width=\columnwidth]{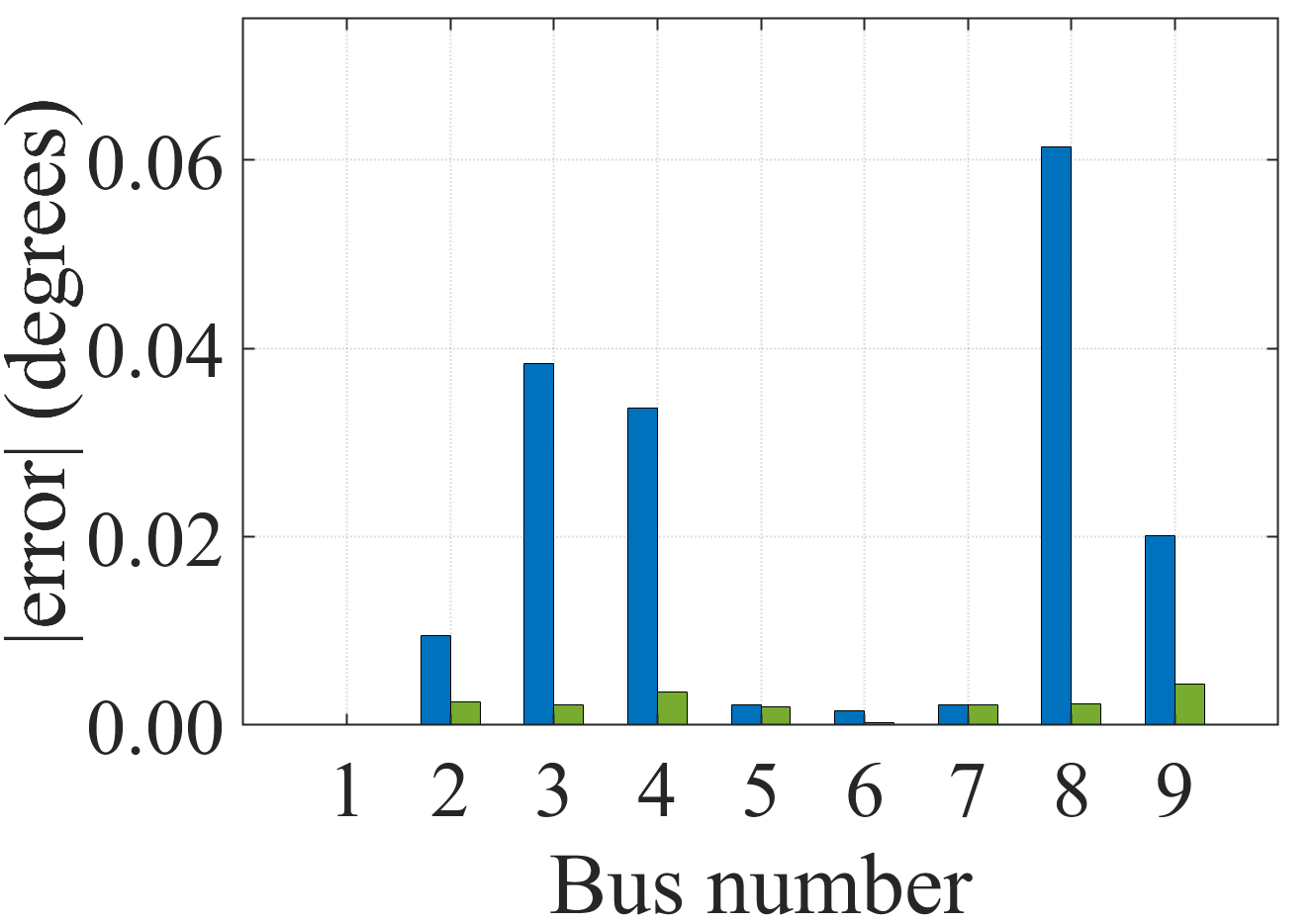}
        \end{minipage}
    }%
    \subfloat[\label{fig:Pi_qc_pp_9bus_pf}]{
        \begin{minipage}{0.24\textwidth}
            \centering
            \includegraphics[width=\columnwidth]{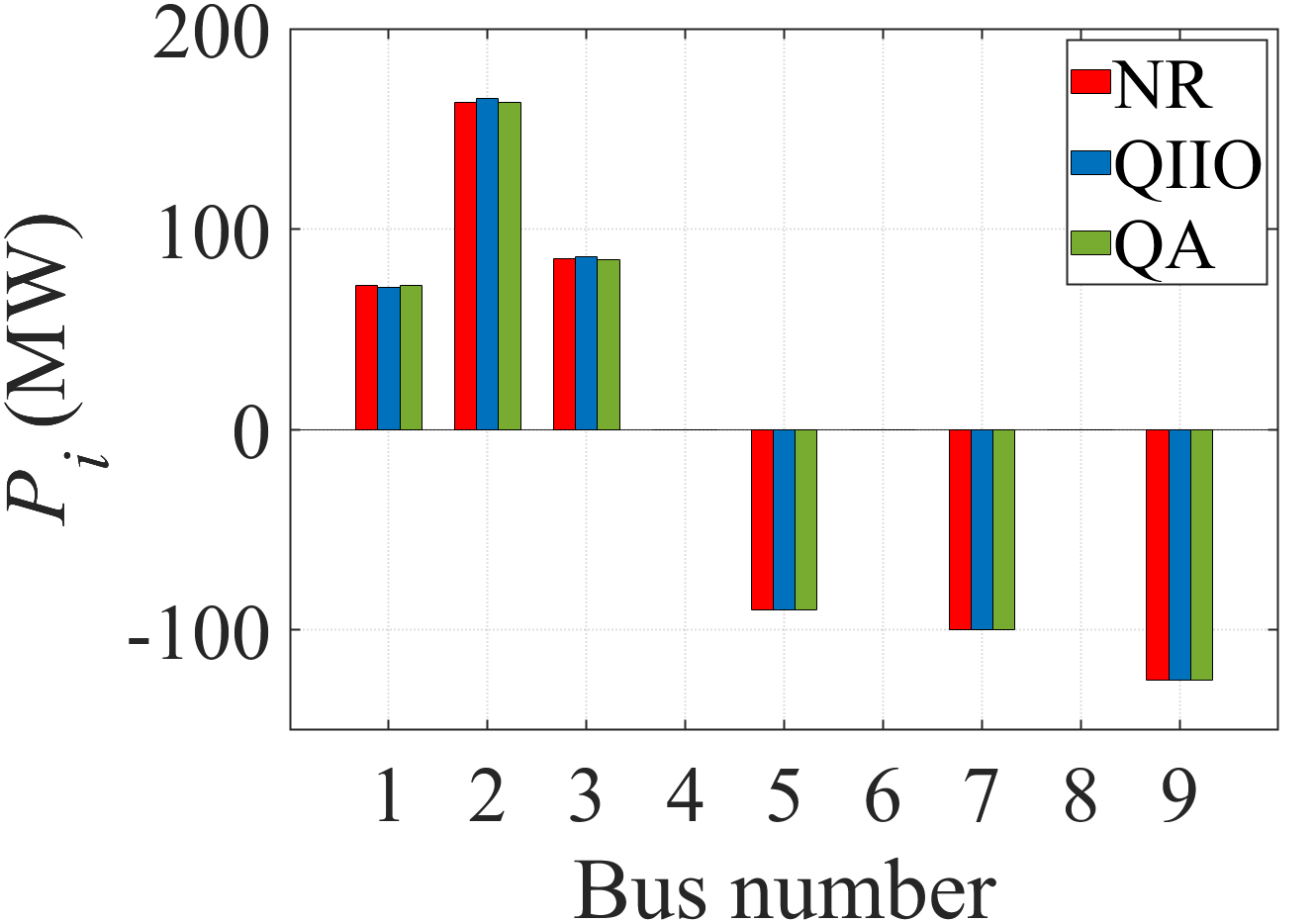} \\
            \includegraphics[width=\columnwidth]{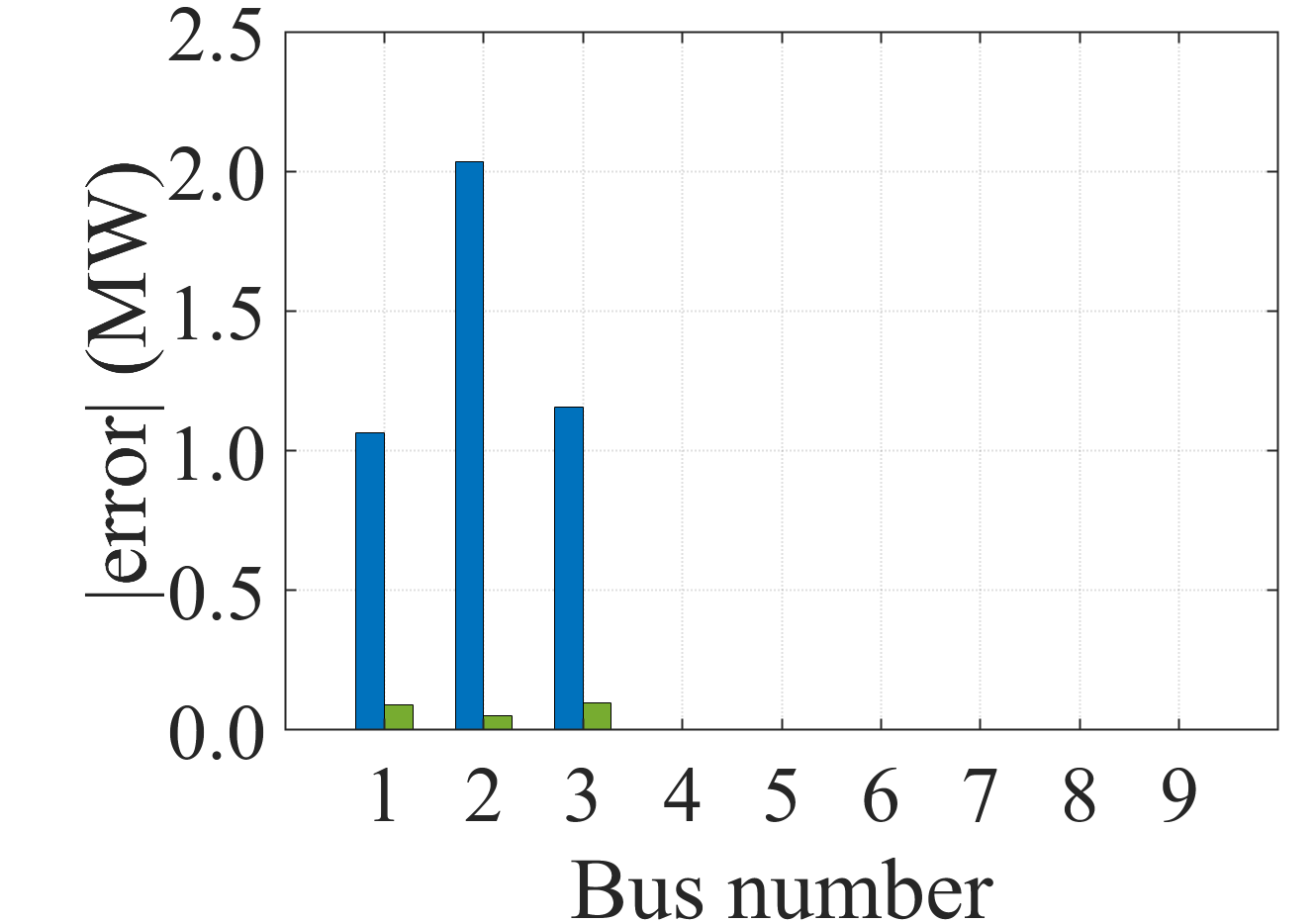}
        \end{minipage}
    }%
    \subfloat[\label{fig:Qi_qc_pp_9bus_pf}]{
        \begin{minipage}{0.24\textwidth}
            \centering
            \includegraphics[width=\columnwidth]{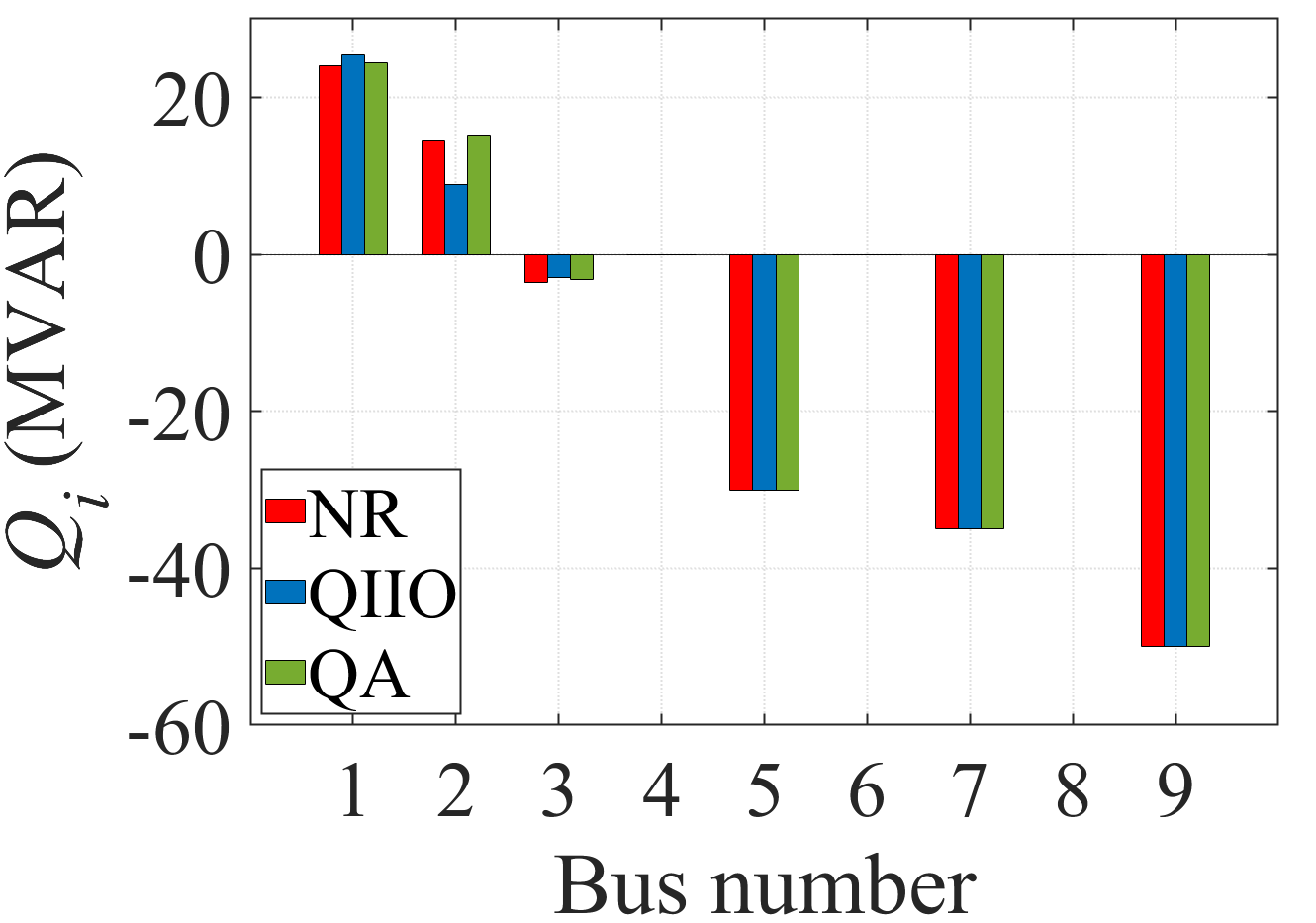} \\
            \includegraphics[width=\columnwidth]{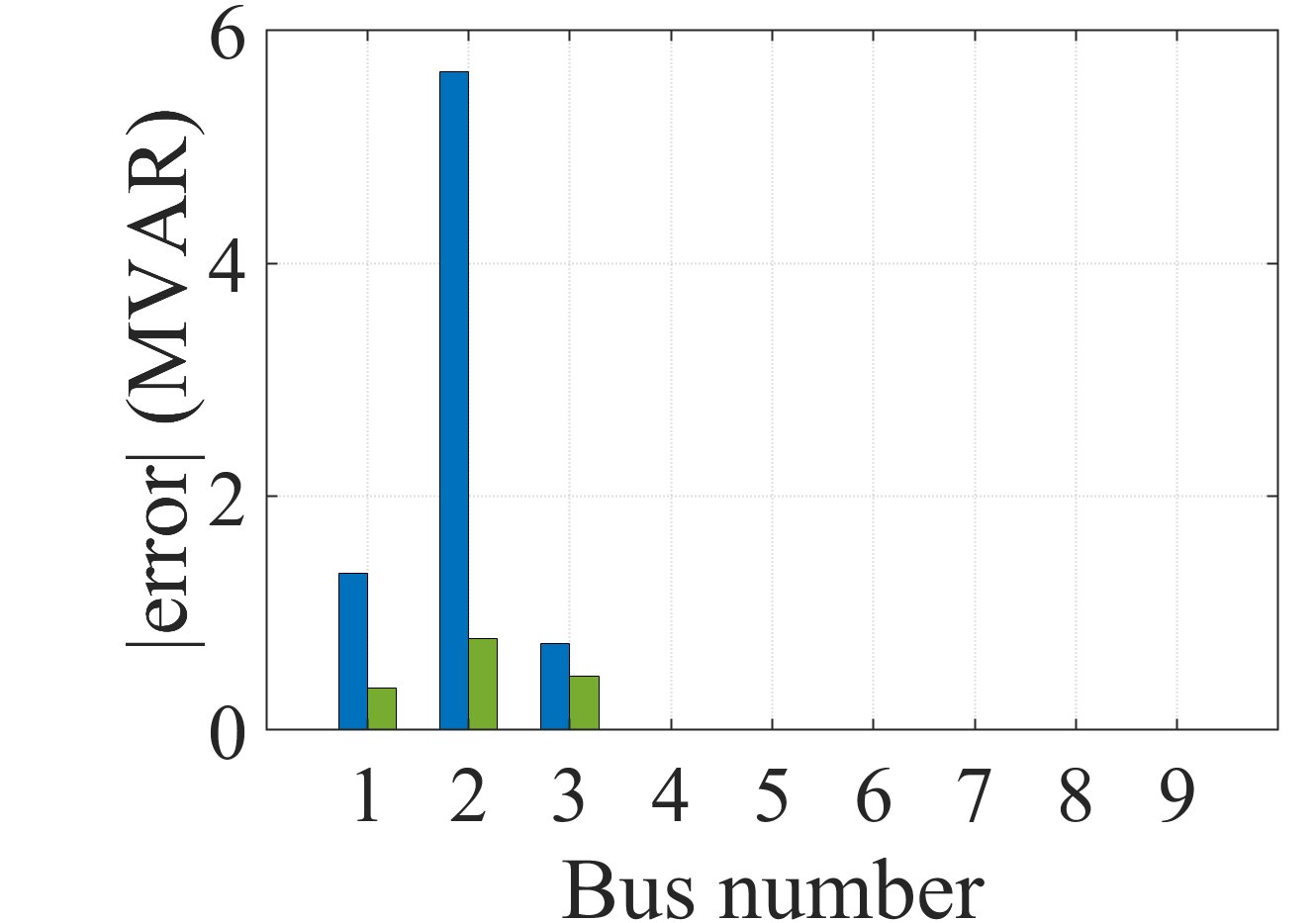}
        \end{minipage}
    }
    \caption{PF results for the IEEE 9-bus test system using QA and QIIO, compared to NR. The top row shows the computed values for (a) $V_i$, (b) $\delta_i$, (c) $P_i$, and (d) $Q_i$. The bottom row shows the absolute errors between QA, QIIO, and NR for each corresponding parameter.}
    \label{fig:qc_pp_9bus_pf}
\end{figure*}

To examine the achieved errors more closely, the PF results for the 9-bus test system obtained using QA and QIIO are compared against that of NR in \cref{fig:qc_pp_9bus_pf}. The top row shows the computed values of $V_i$, $\delta_i$, $P_i$, and $Q_i$, while the bottom row shows the absolute errors of QA, QIIO, relative to NR for each corresponding value. As observed, QA accurately computes both $V_i$ and $\delta_i$, yielding $P_i$ and $Q_i$ values that closely match those of NR. QIIO also produces results within the predefined residual threshold for $V_i$ and $\delta_i$, as shown in \cref{fig:Vi_qc_pp_9bus_pf,fig:deltai_qc_pp_9bus_pf}. However, QA achieves higher accuracy in $P_i$ and $Q_i$ by up to 90\%, as shown in \cref{fig:Pi_qc_pp_9bus_pf,fig:Qi_qc_pp_9bus_pf}. These results indicate that while both QA and QIIO satisfy the overall convergence and accuracy criteria based on average error across all buses, QA demonstrates superior accuracy per bus.

\begin{figure*}[htbp]
    \centering
    \subfloat[\label{fig:Vi_qc_pp_9bus_opf}]{
        \begin{minipage}{0.24\textwidth}
            \centering
            \includegraphics[width=\columnwidth]{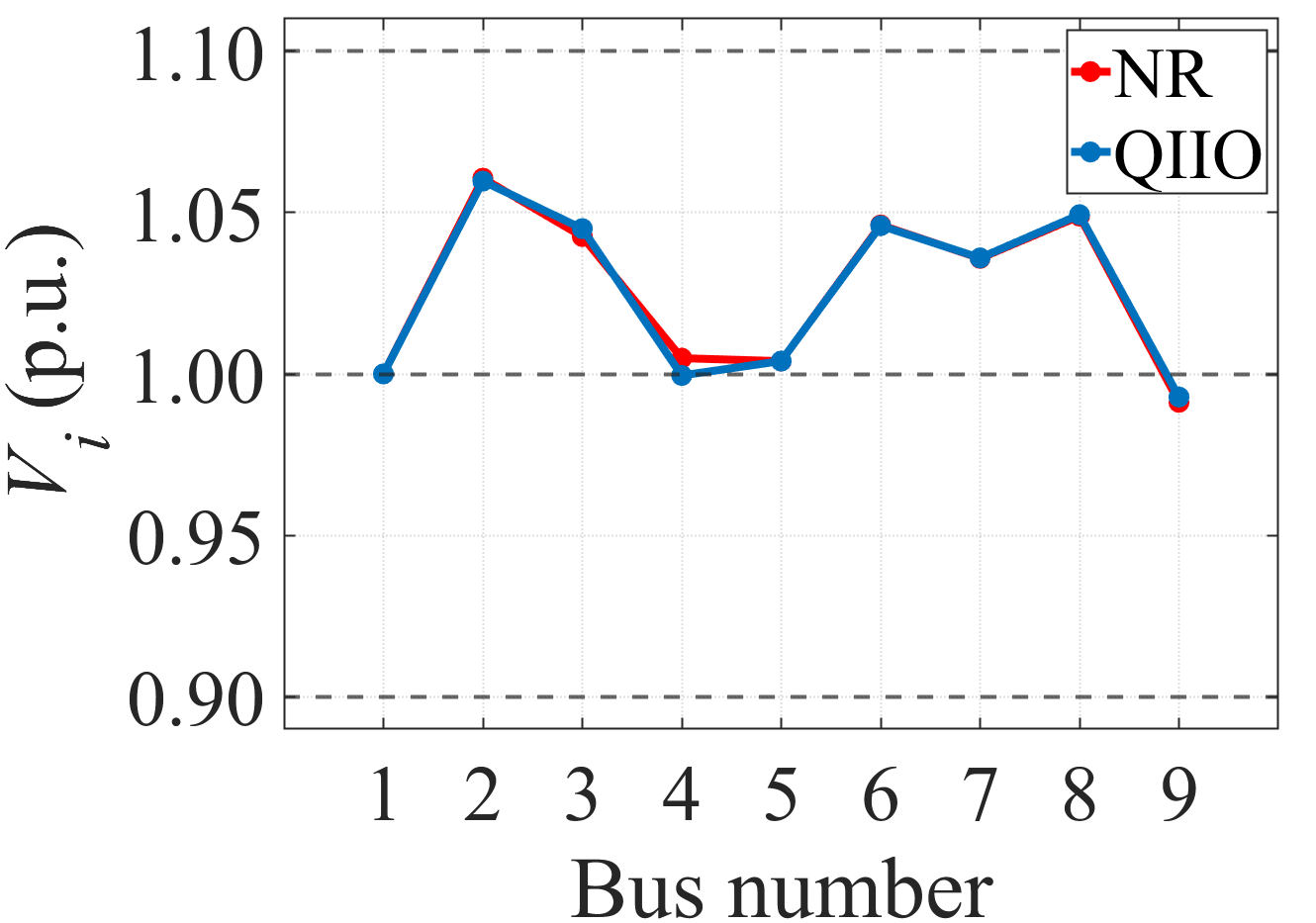} \\
            \includegraphics[width=\columnwidth]{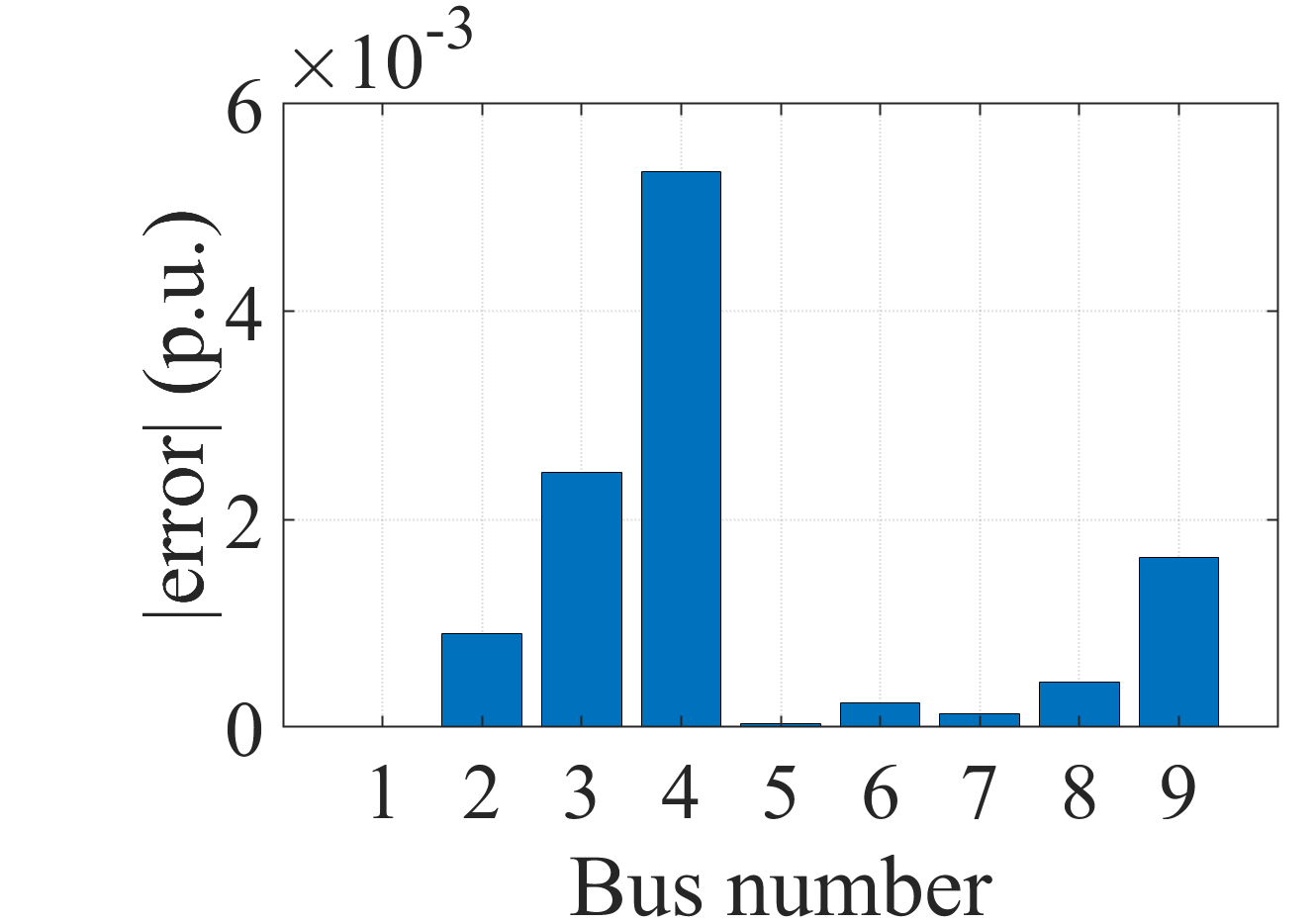}
        \end{minipage}
    }%
    \subfloat[\label{fig:deltai_qc_pp_9bus_opf}]{
        \begin{minipage}{0.24\textwidth}
            \centering
            \includegraphics[width=\columnwidth]{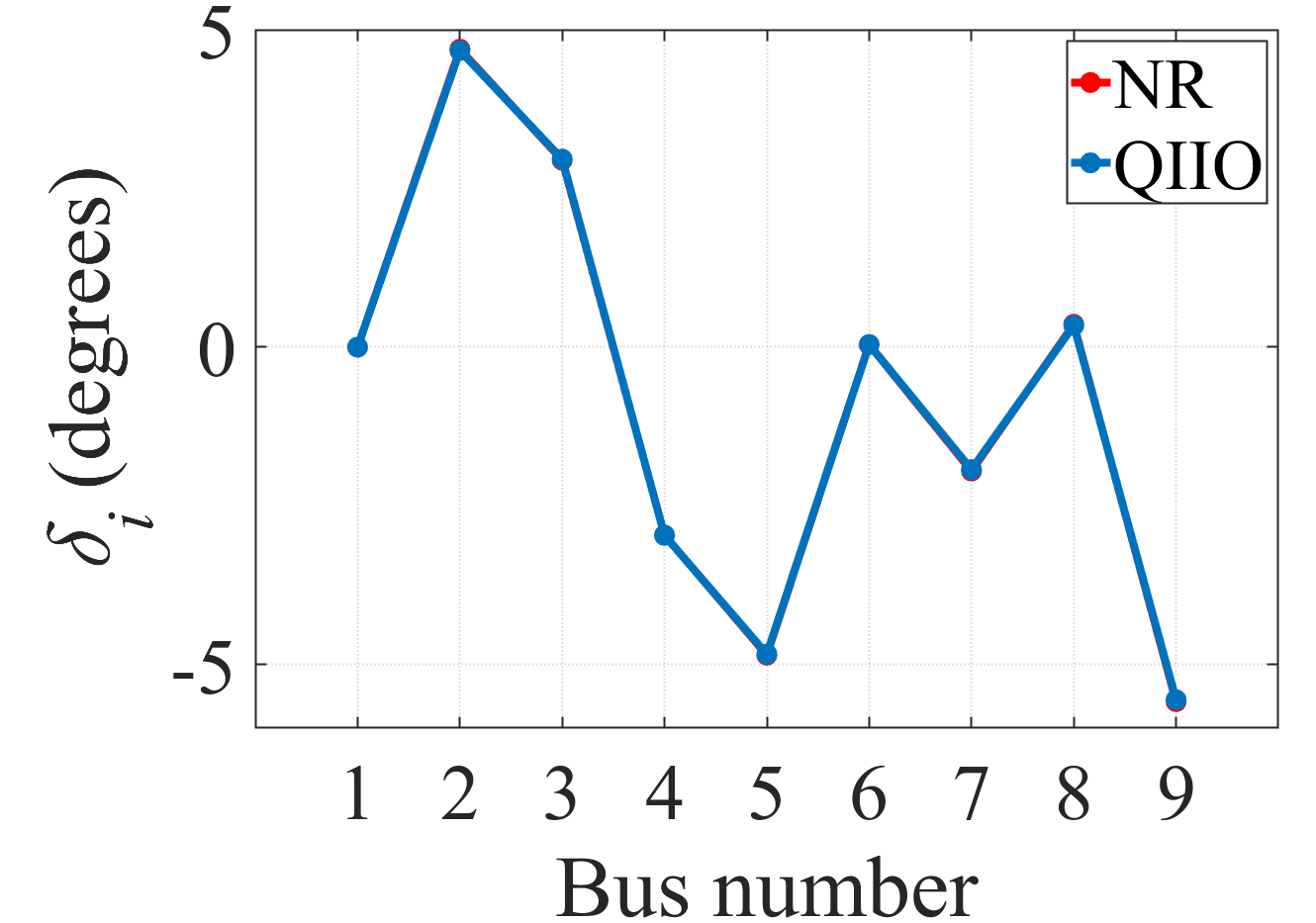} \\
            \includegraphics[width=\columnwidth]{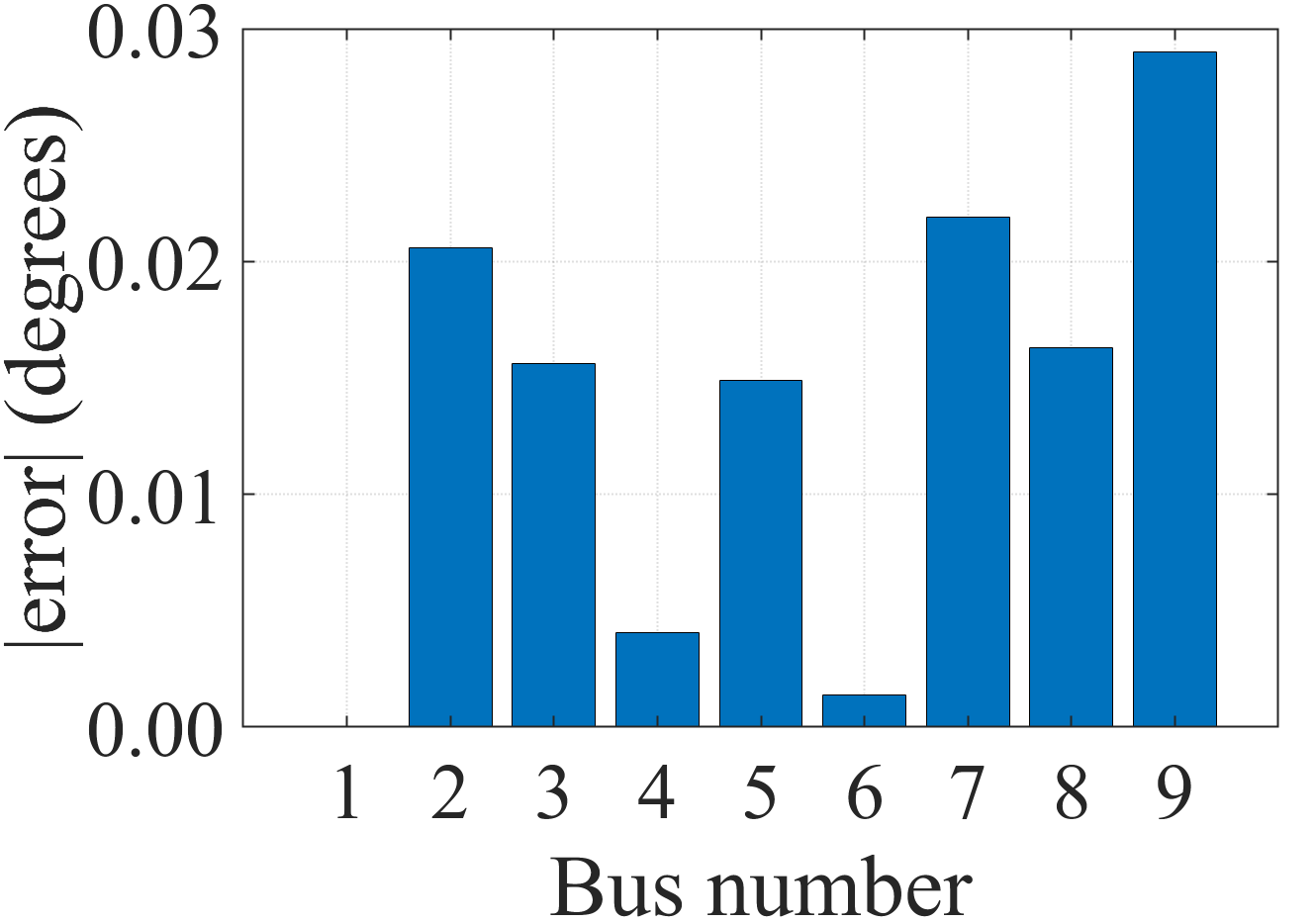}
        \end{minipage}
    }%
    \subfloat[\label{fig:Pi_qc_pp_9bus_opf}]{
        \begin{minipage}{0.24\textwidth}
            \centering
            \includegraphics[width=\columnwidth]{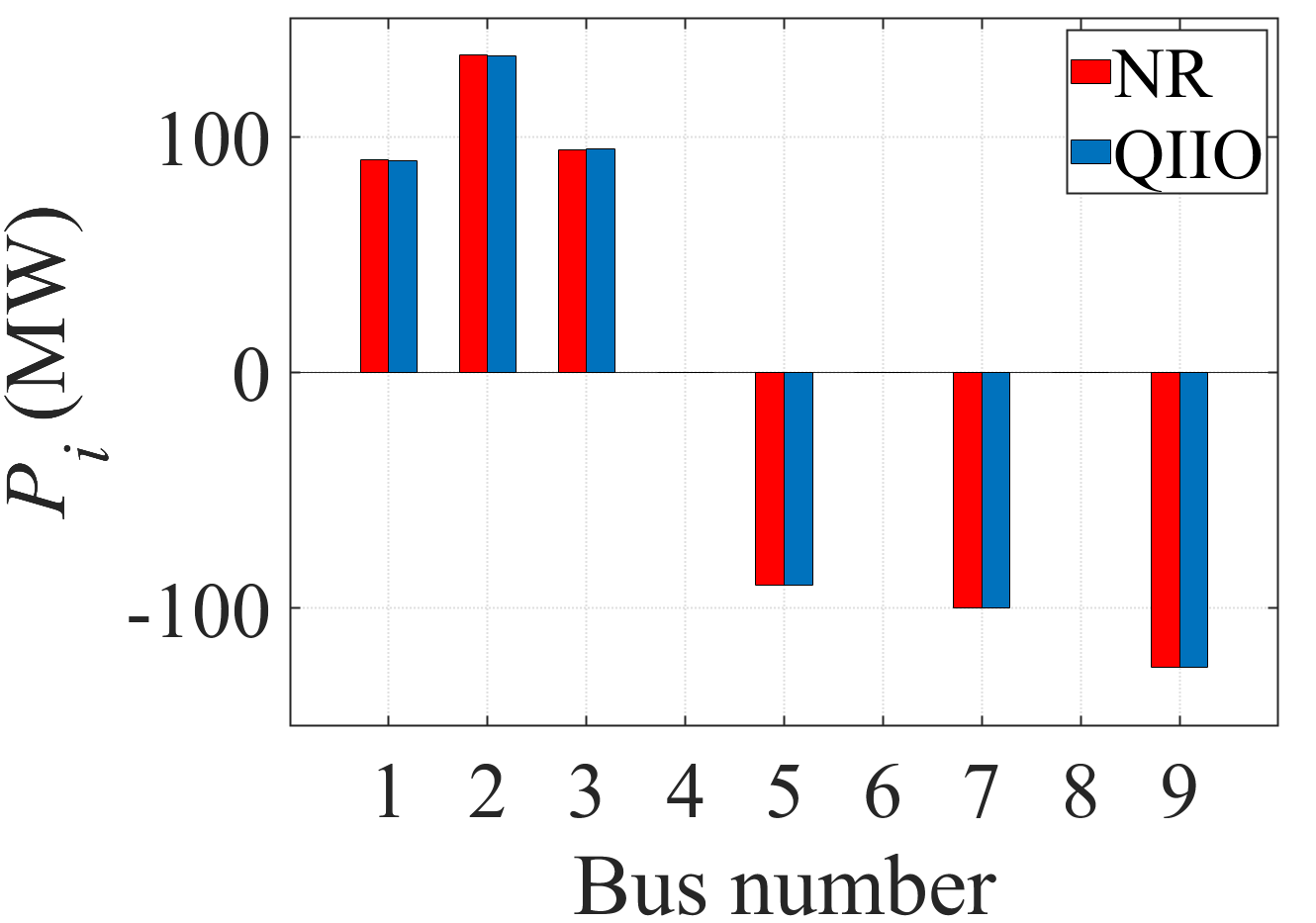} \\
            \includegraphics[width=\columnwidth]{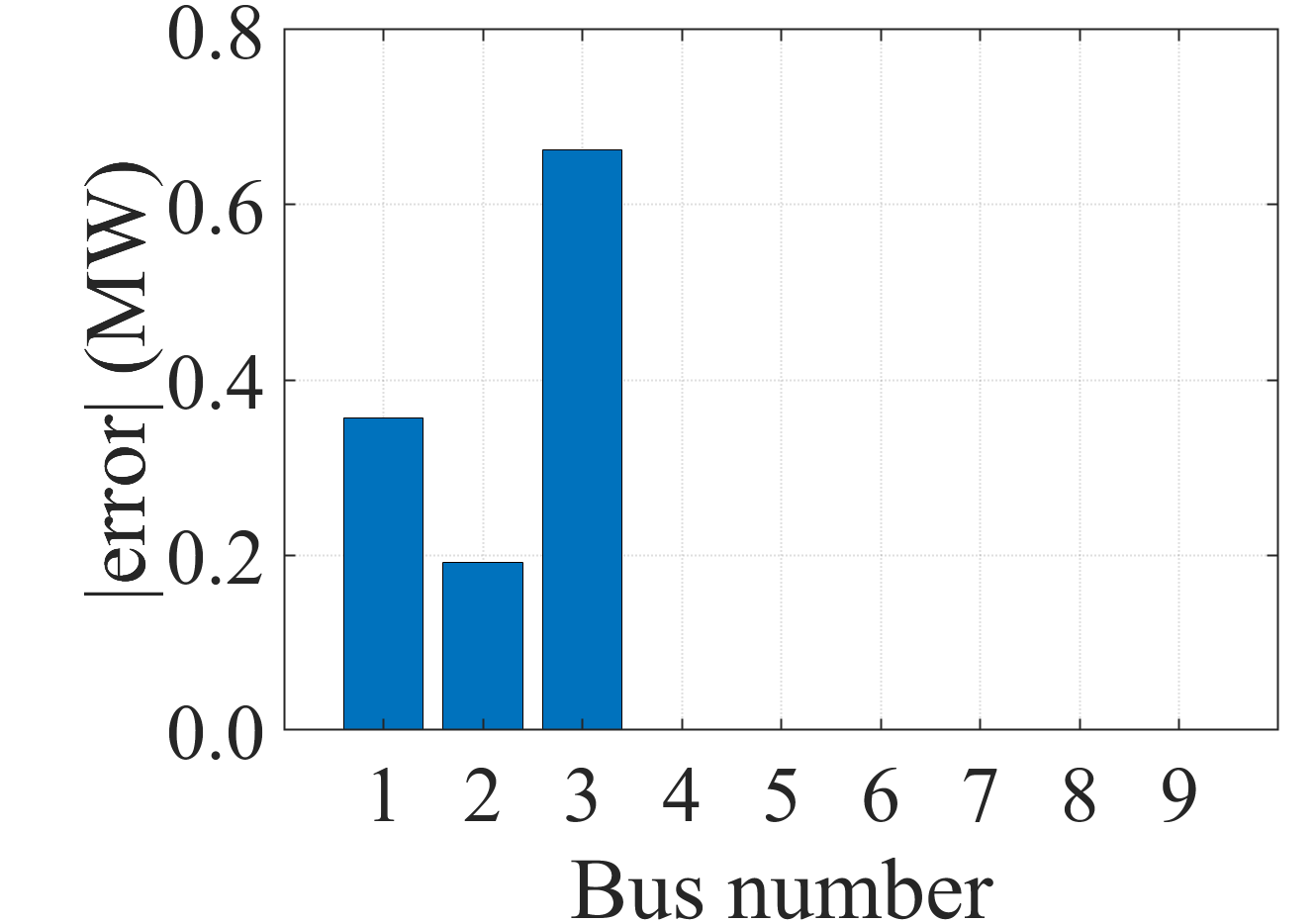}
        \end{minipage}
    }%
    \subfloat[\label{fig:Qi_qc_pp_9bus_opf}]{
        \begin{minipage}{0.24\textwidth}
            \centering
            \includegraphics[width=\columnwidth]{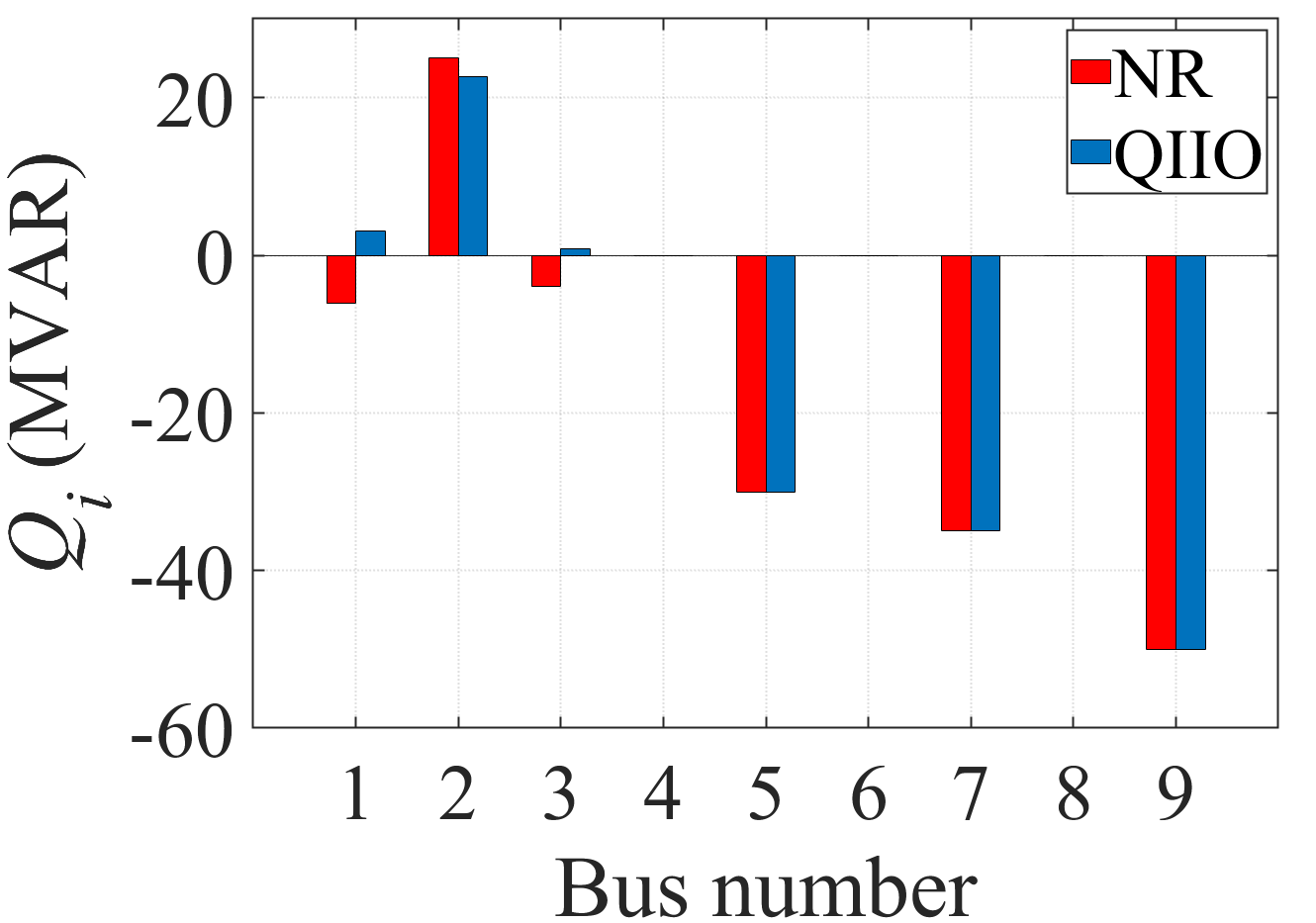} \\
            \includegraphics[width=\columnwidth]{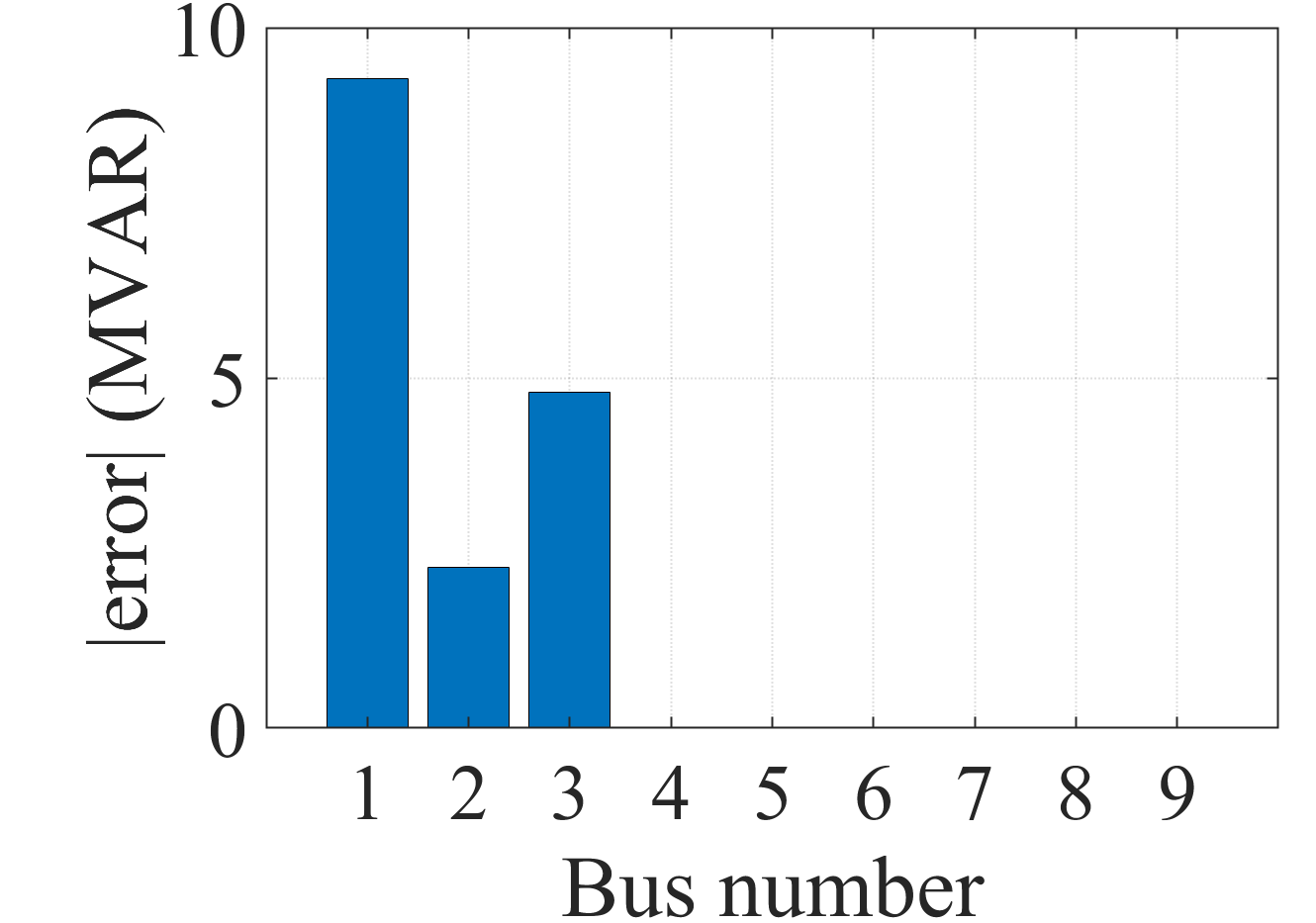}
        \end{minipage}
    }
    \caption{OPF results for the IEEE 9-bus test system using QIIO, compared to NR. The top row shows the computed values for (a) $V_i$, (b) $\delta_i$, (c) $P_i$, and (d) $Q_i$. The bottom row shows the absolute errors between QIIO and NR for each corresponding parameter.}
    \label{fig:qc_pp_9bus_opf}
\end{figure*}

\begin{figure*}[htbp]
    \centering
    \subfloat[\label{fig:Vi_qc_pp_13bus_opf}]{
        \begin{minipage}{0.24\textwidth}
            \centering
            \includegraphics[width=\columnwidth]{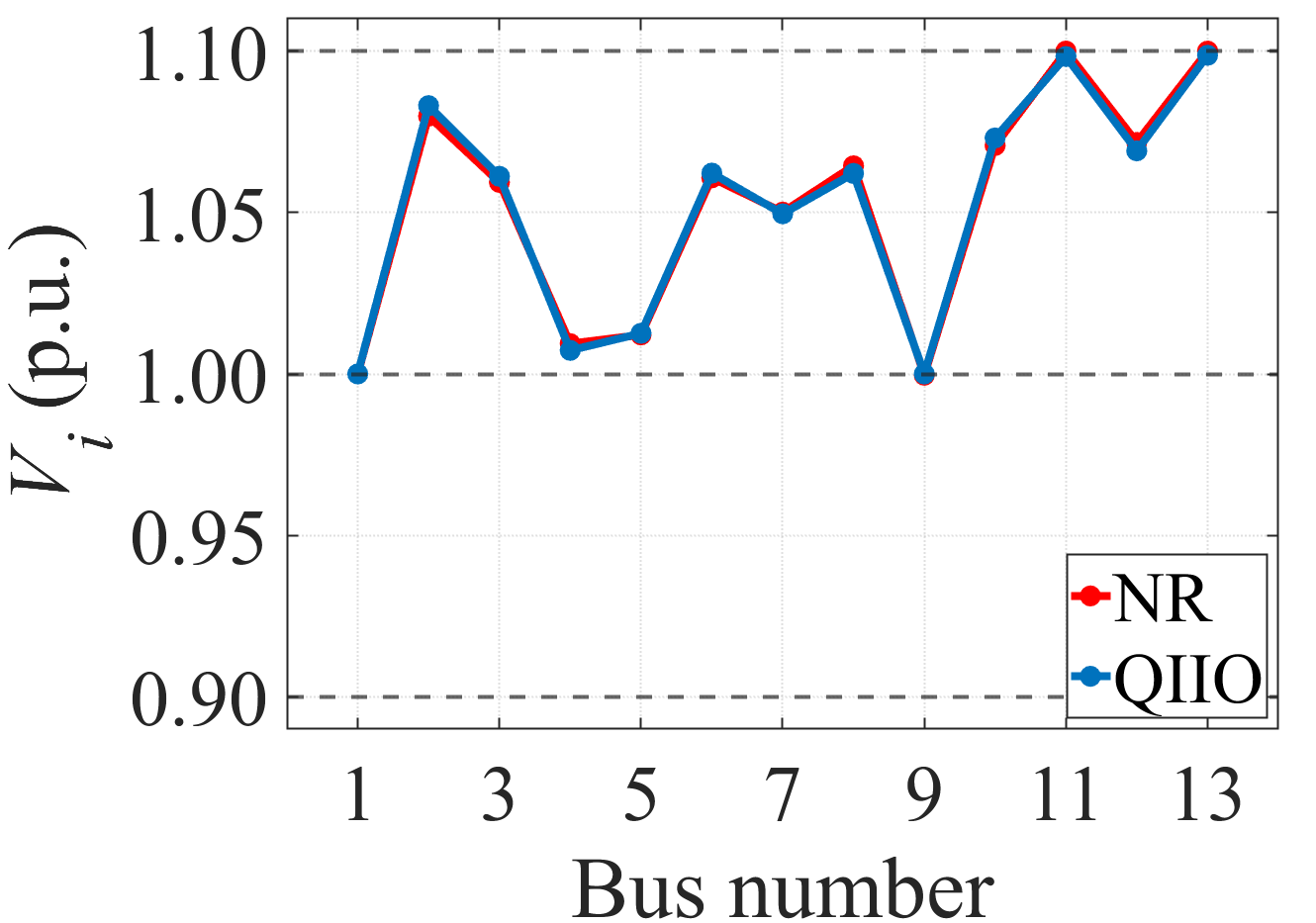} \\
            \includegraphics[width=\columnwidth]{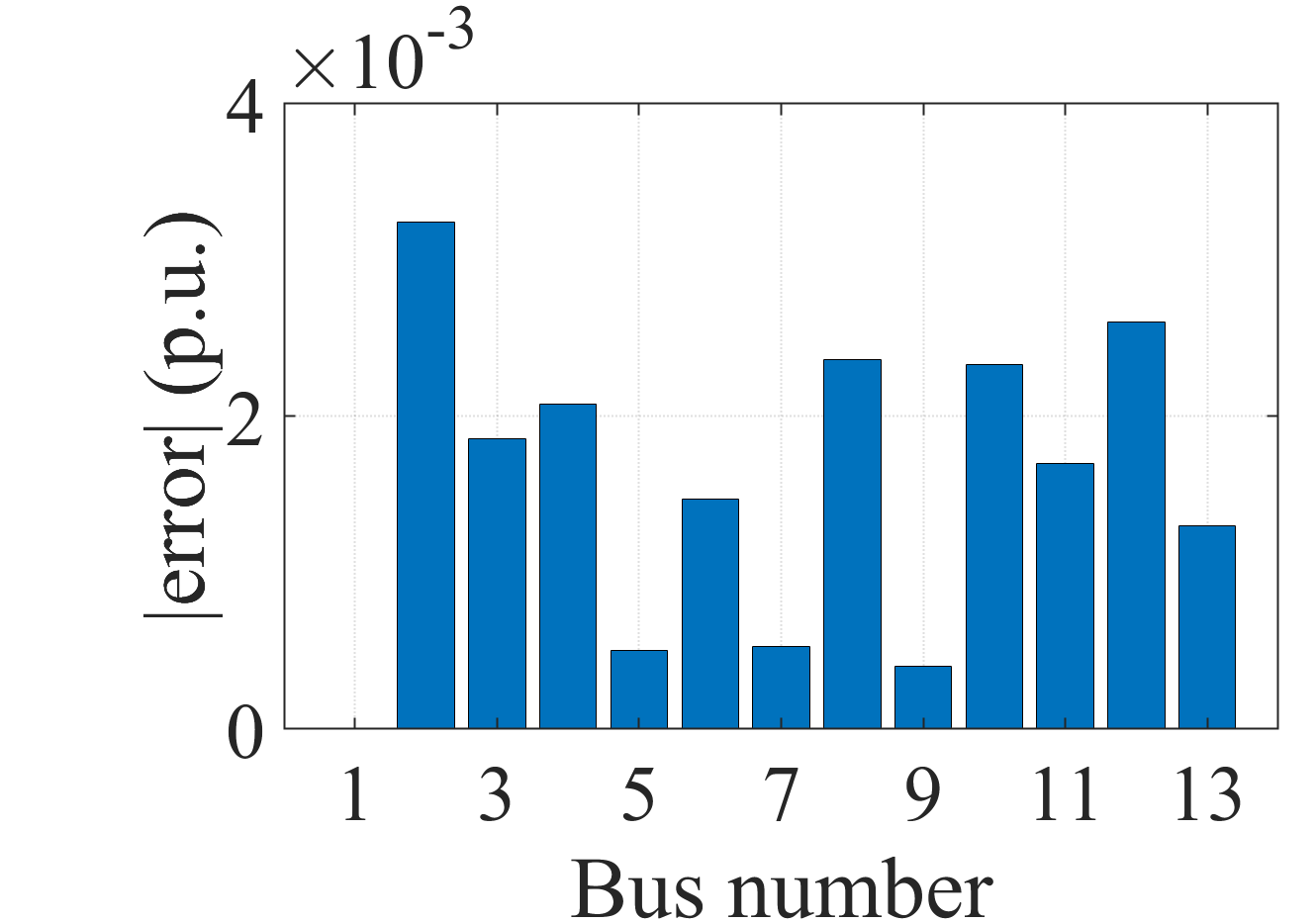}
        \end{minipage}
    }%
    \subfloat[\label{fig:deltai_qc_pp_13bus_opf}]{
        \begin{minipage}{0.24\textwidth}
            \centering
            \includegraphics[width=\columnwidth]{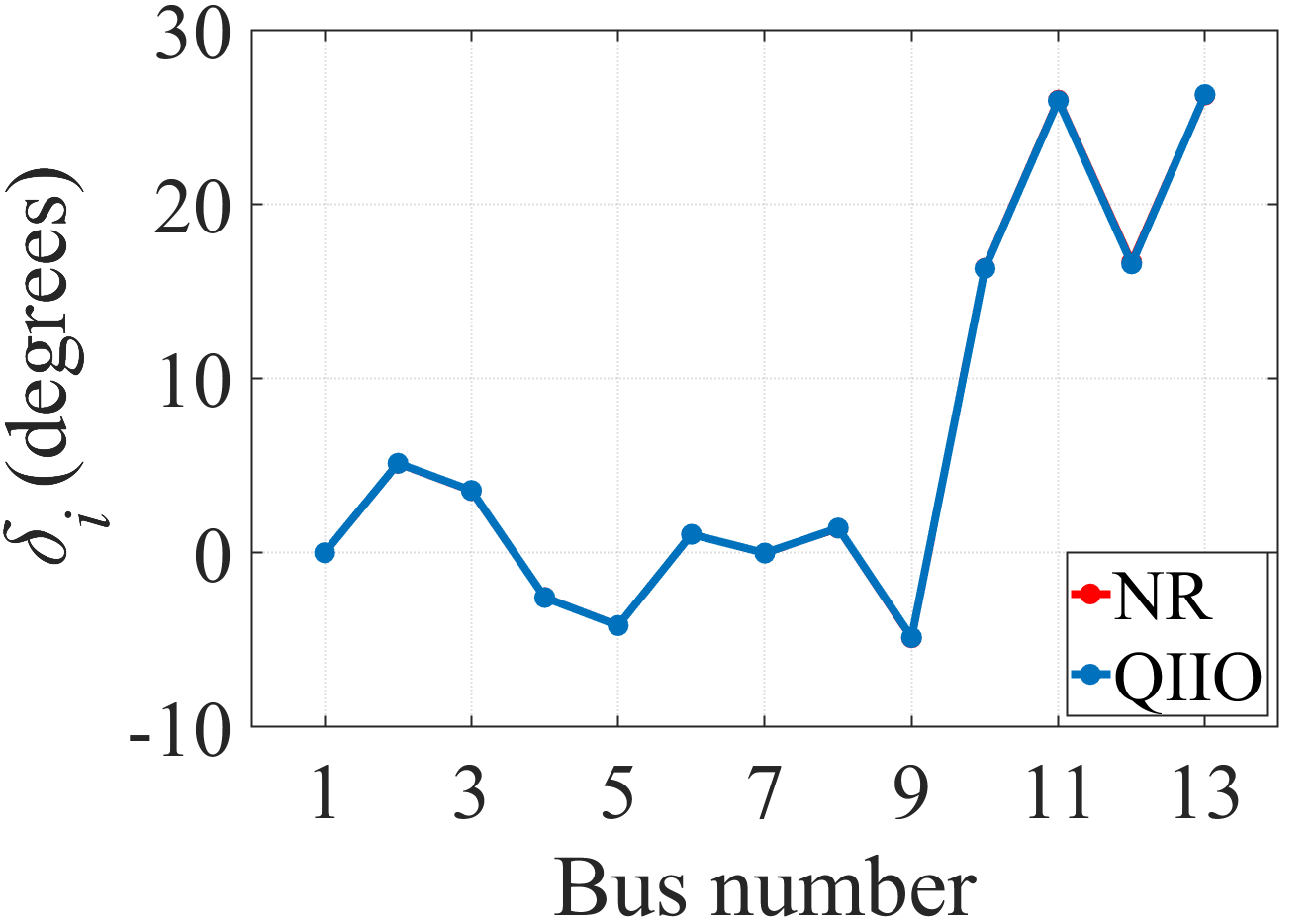} \\
            \includegraphics[width=\columnwidth]{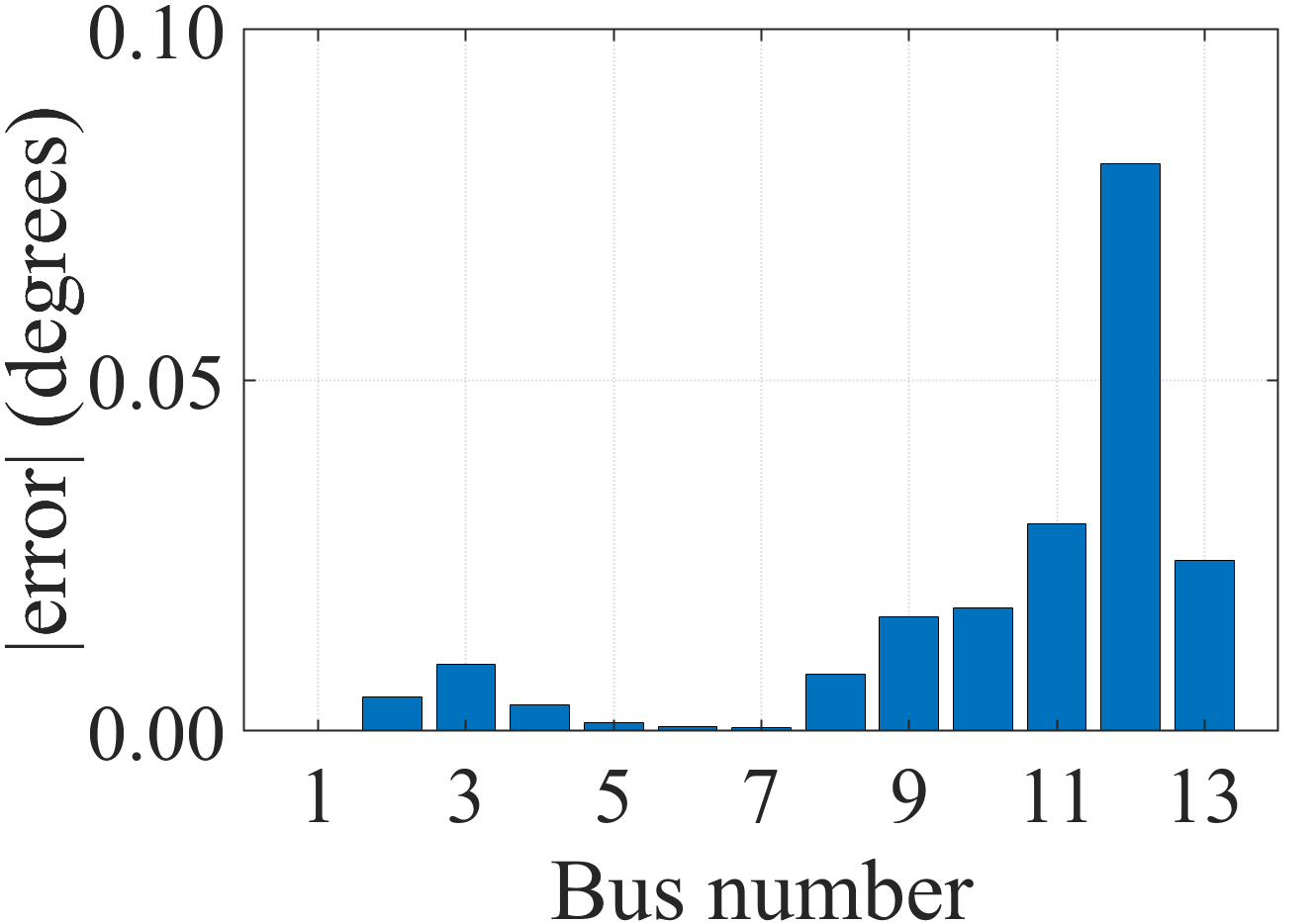}
        \end{minipage}
    }%
    \subfloat[\label{fig:Pi_qc_pp_13bus_opf}]{
        \begin{minipage}{0.24\textwidth}
            \centering
            \includegraphics[width=\columnwidth]{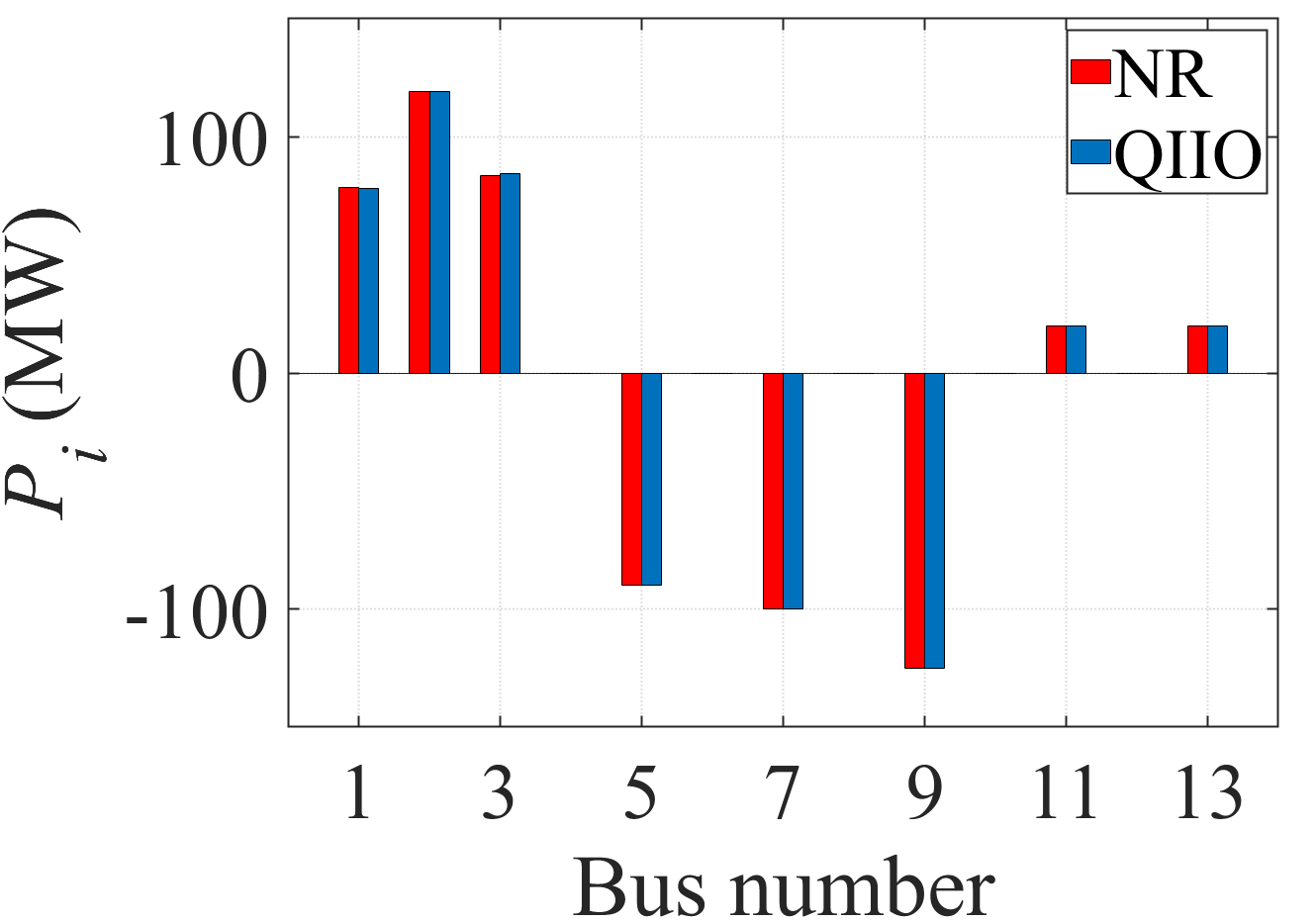} \\
            \includegraphics[width=\columnwidth]{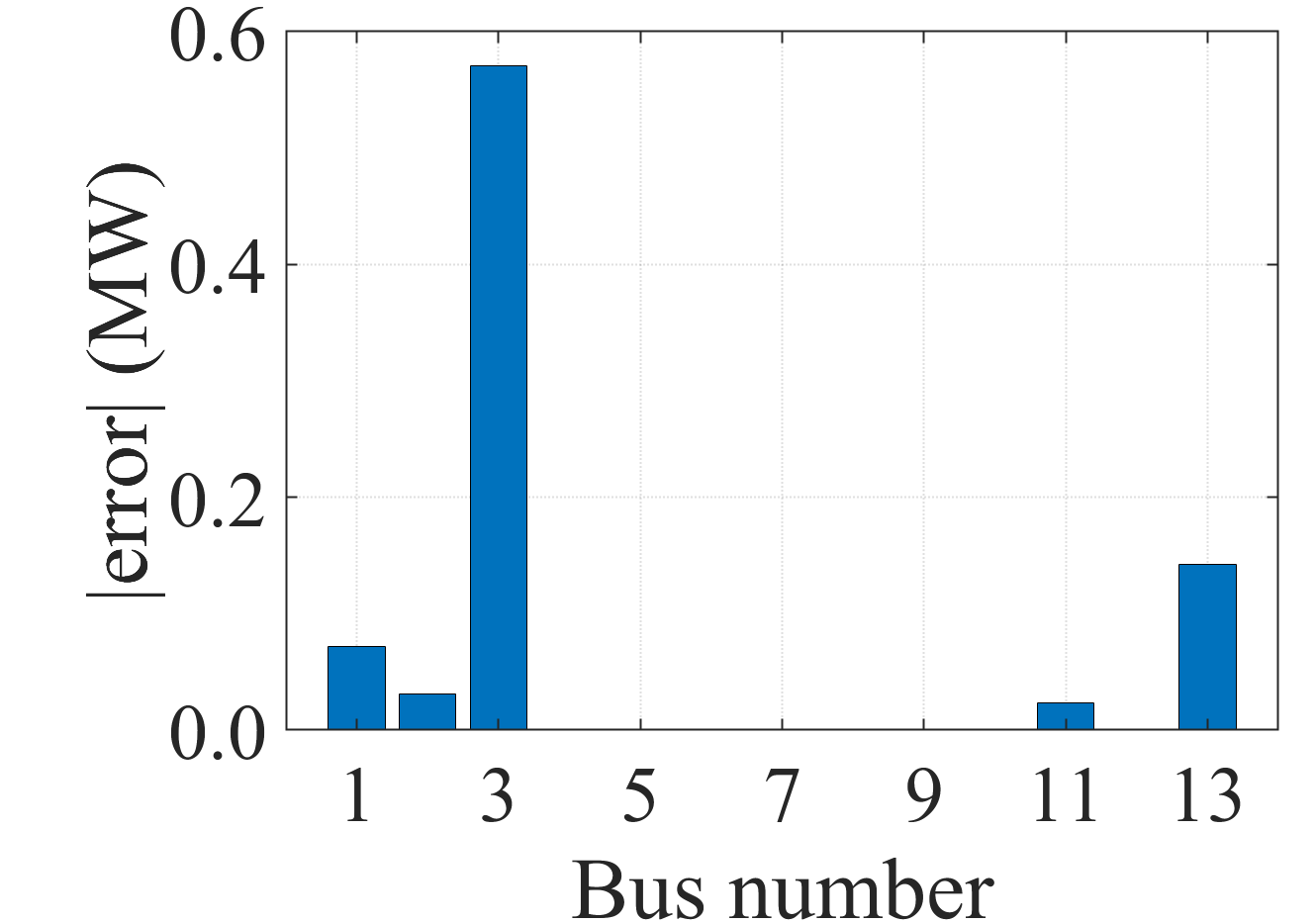}
        \end{minipage}
    }%
    \subfloat[\label{fig:Qi_qc_pp_13bus_opf}]{
        \begin{minipage}{0.24\textwidth}
            \centering
            \includegraphics[width=\columnwidth]{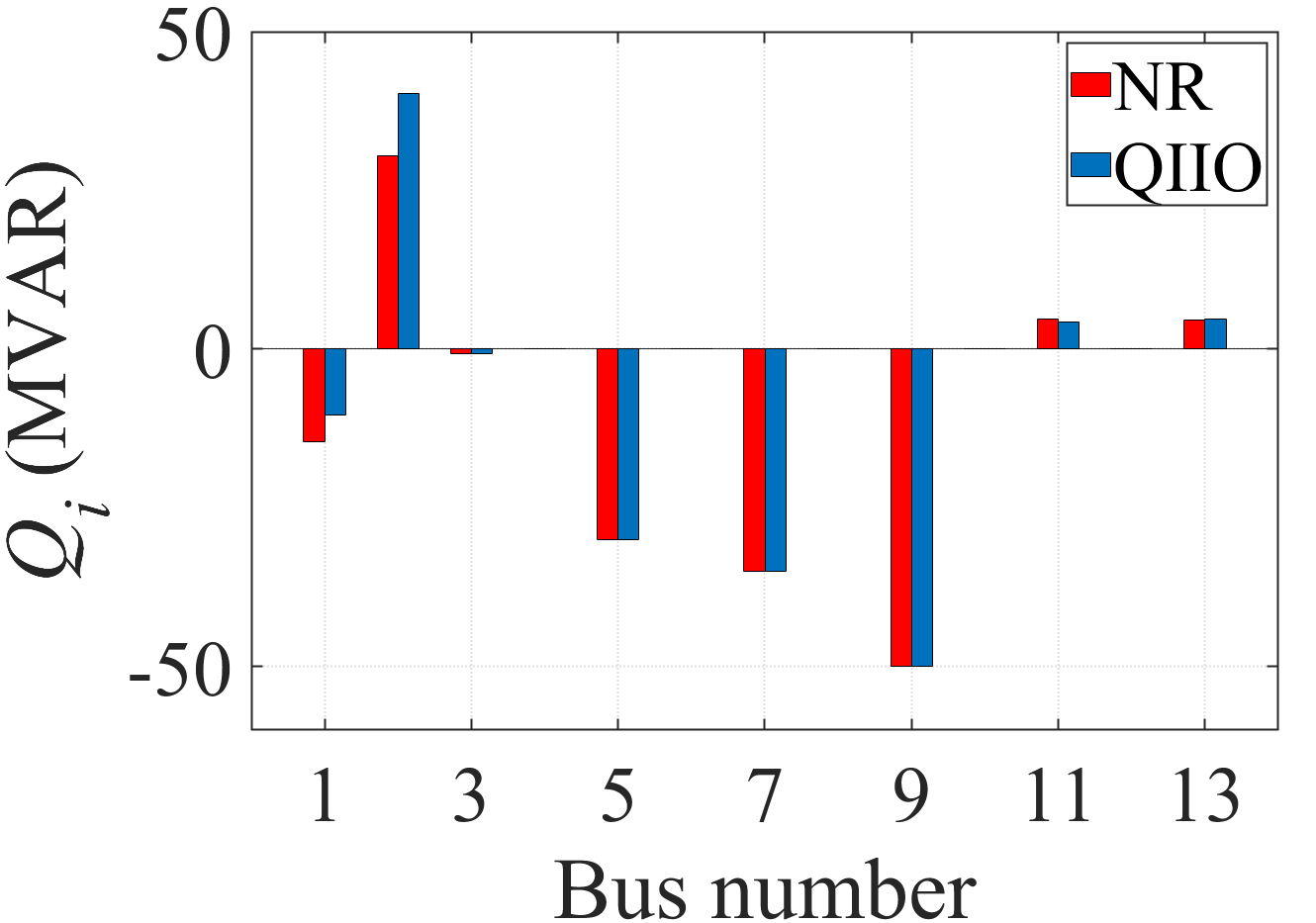} \\
            \includegraphics[width=\columnwidth]{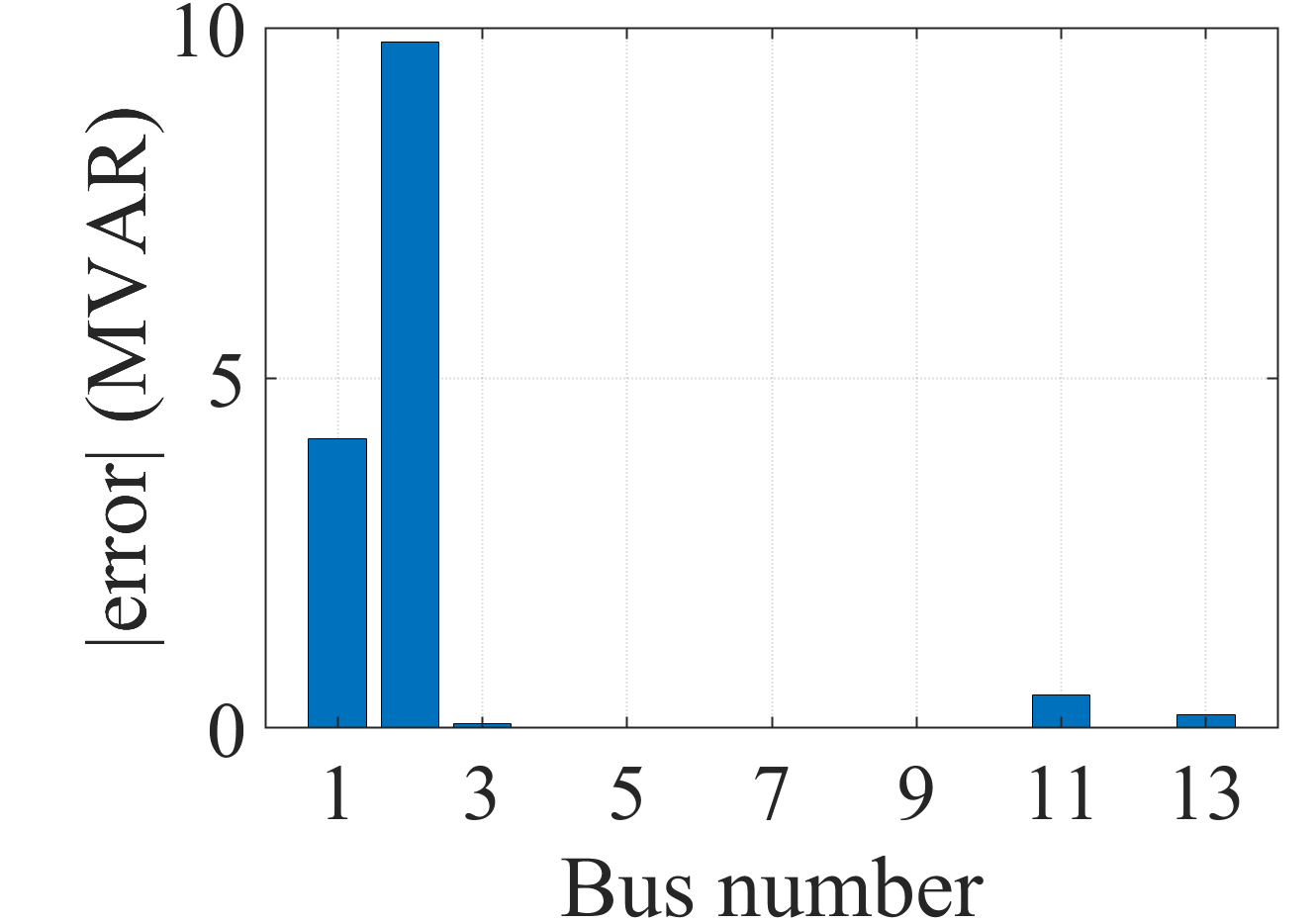}
        \end{minipage}
    }
    \caption{OPF results for the IEEE 13-bus test system with integrated RES using QIIO, compared to NR. The top row shows the computed values for (a) $V_i$, (b) $\delta_i$, (c) $P_i$, and (d) $Q_i$. The bottom row shows the absolute errors between QIIO and NR for each corresponding parameter.}
    \label{fig:qc_pp_13bus_opf}
\end{figure*}

The experiments are extended to include OPF for the 9-bus and 13-bus test systems, for which the results using QIIO are presented in \cref{fig:qc_pp_9bus_opf,fig:qc_pp_13bus_opf}, respectively. QA is not included as the problem graph could not be embedded onto the available hardware graph. Furthermore, due to limited access to QIIO, simulations are terminated after $it_{\text{max}}=200$ iterations, leading to residuals slightly more than the predefined residual threshold, see \cref{tab:solvers,tab:solvers_results}. Nevertheless, the $V_i$ and $\delta_i$ values computed by QIIO exhibit agreement with NR. For the 9-bus test system, the mean deviation in $P_i$ between QIIO and NR is $2.20 \times 10^{-2}$ MW, while in $Q_i$, the mean deviation is $1.13$ MVAR, as shown in \cref{fig:Pi_qc_pp_9bus_opf,fig:Qi_qc_pp_9bus_opf}. For the 13-bus test system, the corresponding deviations are $6.92 \times 10^{-2}$ MW and $1.15$ MVAR, respectively, as shown in \cref{fig:Pi_qc_pp_13bus_opf,fig:Qi_qc_pp_13bus_opf}.

The maximum error in $\mathbf{P}$ for PF is $0.1\%$ using QA and $2\%$ using QIIO for the 9-bus test system. For OPF, QA results are not included due to hardware embedding limitations. Using QIIO, the maximum error in $\mathbf{P}$ is $0.6\%$ for both the 9-bus and 13-bus test systems. The corresponding maximum errors in $\mathbf{Q}$ are $9.2\%$ and $9.7\%$ for the 9-bus and 13-bus test systems, respectively. The relatively higher deviations observed in $\mathbf{Q}$ can be attributed to several factors. First, the IEEE 9-bus system exhibits a low R/X ratio (see \cref{tab:line_params}), which is characteristic of transmission networks. In such systems, $\mathbf{Q}$ calculations are more sensitive to numerical approximations, and it is well known that classical PF solutions may exhibit larger discrepancies in $\mathbf{Q}$ compared to $\mathbf{P}$ (e.g.,~\cite{Talkington2024}). Second, the higher deviations are primarily observed at generator buses, where differences in modeling assumptions between the NR formulation and the RTDS model, such as generator limits or shunt compensation, can lead to slight voltage mismatches. Due to the strong coupling between voltage magnitude and reactive power, even small deviations in $\mathbf{V}$  can lead to amplified errors in computed $\mathbf{Q}$.

\section{Discussion}
Combinatorial PF and OPF formulations enable compatibility with AQC hardware and a discrete modeling approach that ensures convergence. However, discretization introduces approximations that can reduce numerical accuracy, as shown in~\cref{tab:solvers_results}. The quadratic nature of current AQC hardware necessitates reducing higher-order terms via auxiliary variables, and slack binary variables are also needed to encode hard constraints, both of which increase the dimensionality. In our previous study~\cite{kaseb2024power}, we conducted extensive simulations on small-scale test systems, including ill-conditioned cases, to demonstrate the robustness and accuracy of the proposed algorithms. The present study introduces a novel QHIL framework for power systems with high RES integration, leveraging the computational capabilities of QA and QIIO to address combinatorial PF and OPF formulations. Therefore, comprehensive sensitivity analyses under varying load conditions or high RES uncertainty are beyond the scope of this study and are left for future investigation. In addition, in transmission systems, decoupled PF methods may provide more accurate estimates of reactive power (e.g., \cite{Stott1974}). Therefore, future research can investigate the performance of different classical solvers to identify the most suitable approaches for both transmission and distribution systems.

In the algorithms originally proposed in~\cite{kaseb2024power}, the step sizes are not fixed but are adaptively updated. Specifically, larger step sizes are employed during the initial iterations to accelerate convergence when the obtained values are far from the solution. In contrast, smaller step sizes are used as the algorithm approaches the solution to improve stability and accuracy. This adaptive strategy is governed by the residual norm, in which the step size is reduced when the residual falls below predefined thresholds, indicating proximity to the solution and the need for improved numerical stability. This strategy is consistent with standard practices in gradient-based methods, among others, that start with a large step size and reduce until a sufficient decrease condition is met (e.g., \cite{Dennis1996}). In addition, the step sizes are updated per bus, enhancing numerical stability, mitigating cumulative discretization errors, and reducing the number of hardware calls. For example, oscillating the value obtained at a given bus across successive iterations indicates that the step size is too large and is accordingly reduced to ensure convergence.

Despite current hardware limitations, early-stage investigations are essential to assess the potential of quantum and quantum-inspired technologies for power system applications, as has already been explored in other domains such as computational fluid dynamics (e.g., \cite{Schalkers2024}). Therefore, this study focuses on methodological development and feasibility rather than large-scale deployment, which is not yet practically attainable with existing quantum hardware. Ongoing hardware advancements, particularly those enabling many-body interactions, higher qubit counts, and improved connectivity, are expected to alleviate issues related to problem size. Yet, we have explored several strategies to mitigate these challenges within current hardware limitations. In related studies~\cite{kaseb2024power}, we demonstrate that reformulating the problem using spin variables $\{\pm 1\}$ instead of binary variables $\{0,1\}$ can reduce the number of decision variables and couplers needed. In addition, a partitioned formulation is proposed that solves subsets of the problem and decreases the problem size per iteration without compromising accuracy. However, developing problem-specific schemes for more efficient reduction of higher-order terms remains an important direction for future study.

\section{Conclusion} \label{sec:conclusion}
This study is motivated by the increasing complexity of renewable-integrated power systems and the emerging potential of AQC to address large-scale nonlinear optimization problems. We develop a QHIL framework to bridge the gap between real-time power system modeling and control and AQC hardware. We also develop a middleware that ensures efficient real-time communication between QIIO and RTDS\textsuperscript{\textregistered}. The framework is validated using the IEEE 9-bus test system and a modified version incorporating solar and wind farms. The algorithms developed in \cite{kaseb2024power} are used to solve the combinatorial PF and OPF formulations. Based on the observations, lessons learned, and results, we demonstrate the feasibility of the proposed QHIL framework and identify practical challenges associated with current and near- to mid-future hardware capabilities.

The tolerance threshold of $10^{-2}$ and $it_{\text{max}}=200$ are selected as stopping criteria for PF and OPF due to limited hardware access and the computational cost associated with QA and QIIO. Within this relatively relaxed tolerance, the results show good agreement with those obtained using the NR solver in Pandapower. Furthermore, we observe that QIIO exhibits better scalability in terms of computational processing time compared to the NR solver. As the problem size increases, the growth rate of the computational processing time of QIIO remains lower than that of NR.

While the average and per bus deviations in $\mathbf{P}$ remain below 1\% for all experiments, the corresponding deviations in $\mathbf{Q}$ are comparatively higher. This behavior is consistent with established findings showing that voltages are generally more sensitive to variations in $\mathbf{Q}$ than in $\mathbf{P}$ (e.g.,~\cite{Talkington2024}). Consequently, small deviations in $\mathbf{V}$ can lead to amplified errors in computed $\mathbf{Q}$. Furthermore, we observe that while QA and QIIO satisfy the overall accuracy criteria based on average error across all buses, QA demonstrates superior accuracy per bus. In this perspective, a bus-specific residual, rather than relying on average deviations, can potentially enhance accuracy per bus.

For the current AQC hardware, which is not yet as readily accessible or scalable as mature classical hardware, computational overhead remains a key limitation for real-time applications. Nevertheless, the information flow within the proposed QHIL framework operates effectively and demonstrates reliable communication between components. That is, once fault-tolerant quantum hardware becomes available, the framework is expected to respond efficiently to dynamic changes in system topology or operating conditions in real time. This capability arises from the algorithm’s iterative structure and warm-starting strategy. When system conditions change, the solution from the previous run is used as the initial point for the next run, which enables rapid re-convergence. Note that this dynamic adaptability is independent of the residual threshold, which only determines the stopping criterion of each individual run. 

Furthermore, current QA is constrained by limited qubit counts, sparse connectivity, and restricted precision in coupling coefficients, which collectively hinder its application to large-scale problems. These challenges are expected to diminish as quantum hardware matures. In contrast, QIIO can handle up to 100,000 decision variables and therefore does not currently face the same scalability limitations. Another key observation is that middleware design and data exchange efficiency are highly platform-specific. That said, different hardware platforms require specific middleware implementations even when solving the same PF and OPF formulations. This observation highlights the importance of developing standardized, adaptable middleware for future real-time applications across various hardware platforms.

\section*{Appendix}

\begin{figure}[t!]
    \centering
    \includegraphics[width=0.9\columnwidth]{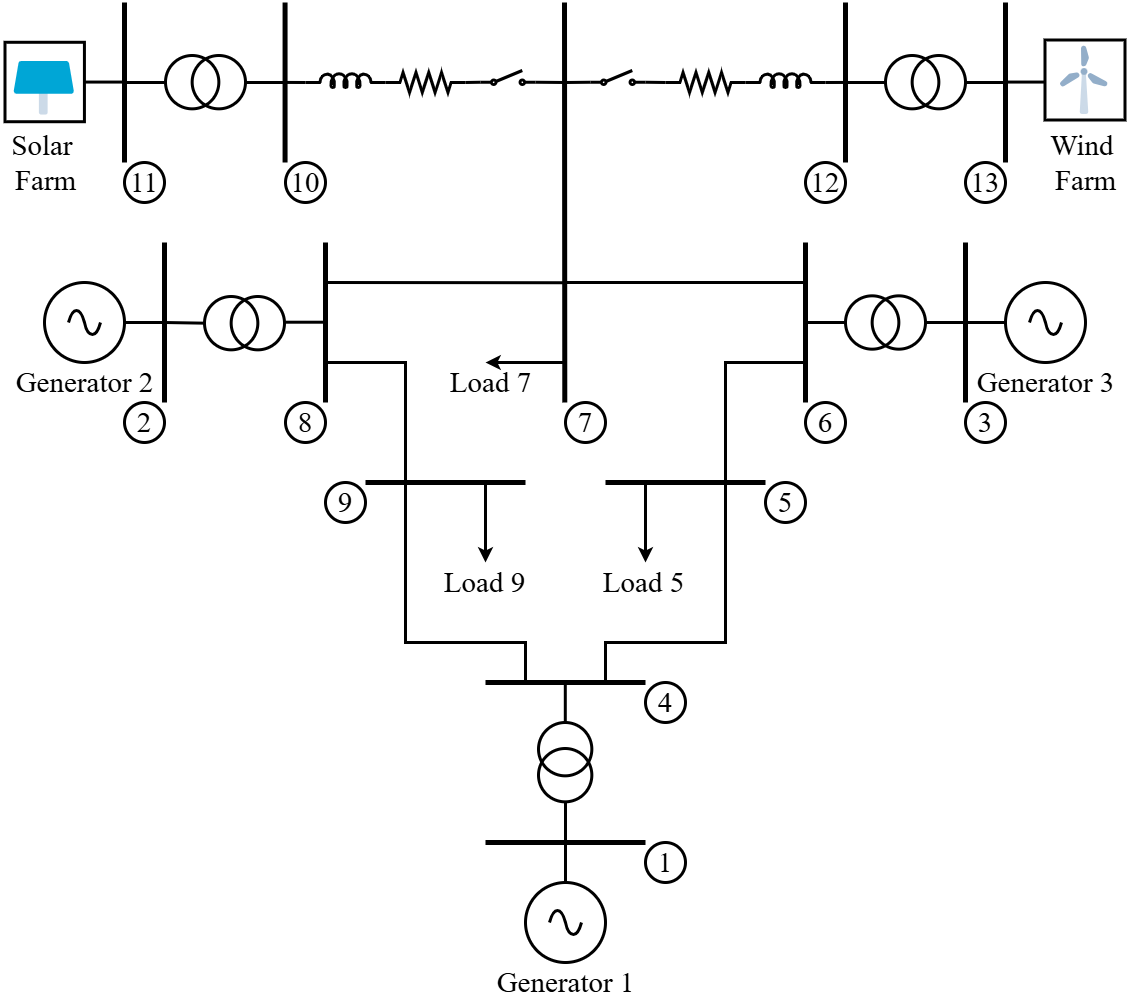}
    \caption{The IEEE 9-bus test system representation with integrated RES. The system includes nine buses, three generators at buses 1, 2, and 3 (\emph{PV} buses), and loads at buses 5, 7, and 9 (\emph{PQ} buses). A 1 MW solar farm and a 2.5 MW wind farm are connected to bus 7, which are scaled to 20 MW each. Both RES units are connected using step-up transformers and \emph{RL} branches, maintaining a constant short circuit ratio. Voltage controllers at the RES buses allow them to function as \emph{PV} buses in PF and OPF formulations.}
    \label{fig:case9_pv_wind}
\end{figure}

The IEEE 9-bus test system is a simplified representation of a power grid, with nine buses, three generators, three transformers, and six transmission lines, as depicted in \cref{fig:case9_pv_wind}. Three of the buses are categorized as \emph{PV} buses, indicating that they are equipped with generators that provide active power and regulate bus voltage by controlling reactive power. These generators are connected to buses 1, 2, and 3. The remaining six buses are \emph{PQ} buses, which may or may not be connected to a load. Specifically, loads are connected to buses 5, 7, and 9. Three step-up transformers connect the generators to the transmission network, and six transmission lines interconnect the buses.

A 1 MW solar farm and a 2.5 MW wind farm have been modeled and connected to bus 7 to integrate RES, see \cref{fig:case9_pv_wind}. These farms, Generators 4 and 5 in \cref{tab:gen_params}, are then scaled up to 20 MW each using the sub-step scaling option available when modeling in RTDS\textsuperscript{\textregistered}. Each RES is connected to the power system via a step-up transformer and a series \emph{RL} branch, ensuring that the farms' short-circuit ratio remains constant as the power output varies. A switch is also included to allow for the connection and disconnection of the RES when simulating various operational conditions. Since these RES units are equipped with voltage controllers in the grid-side converters at their respective connecting buses, they can be effectively modeled as \emph{PV} buses in the PF and OPF formulations.

The power output of each RES is measured at the grid-side converter, corresponding to the point of common coupling (PCC), such that the measured active power represents the net power injected into the grid. Therefore, converter losses are inherently excluded from the measured values. Consequently, the maximum active power reported in \cref{tab:gen_params} (19.8 MW) corresponds to the actual power delivered to the grid, which is slightly lower than the nominal 20 MW rating. The cost parameters for these RES units are set lower than those for conventional generators, reflecting their lower operational and maintenance costs while providing a realistic economic representation in the OPF simulations. \cref{tab:gen_params,tab:load_params,tab:line_params,tab:gencost_params} respectively summarize the test system's load, line, and cost parameters, considering base values of 345 kV and 100 MVA.

\begin{table}[t!]
    \centering
    \caption{Generator parameters for the IEEE 9-bus test system with integrated RES, including generator locations, capacities, and operational limits.}
    \label{tab:gen_params}
    \resizebox{\columnwidth}{!}{
    \begin{tabular}{ccccccccc}
        \hline
        Sr. & Bus & $P_i^G$ & $V_i$ & $\delta_i$ & $\overline{P^G_i}$ & $\underline{P^G_i}$ & $\overline{Q^G_i}$ & $\underline{Q^G_i}$ \\
        No. & $i$ & (MW) & (p.u.) & (deg.) & (MW) & (MW) & (MVAR) & (MVAR) \\
        \hline
        1 & 1 & - & 1 & 0 & 250 & 10 & 300 & -300 \\
        2 & 2 & 163 & 1 & - & 300 & 10 & 300 & -300 \\
        3 & 3 & 85 & 1 & - & 270 & 10 & 300 & -300 \\
        4 & 11 & 19.8 & 1 & - & 19.8 & 19.8 & 20 & -20 \\
        5 & 13 & 19.8 & 1 & - & 19.8 & 19.8 & 20 & -20 \\
        \hline
    \end{tabular}
    }
\end{table}

\begin{table}[t!]
    \centering
    \caption{Load parameters for the IEEE 9-bus test system with integrated RES, containing the active and reactive power demands at each load-connected bus.}
    \label{tab:load_params}
    \begin{tabular}{cccc}
        \hline
        Sr. & Bus & $P_i^D$ & $Q_i^D$ \\
        No. & $i$ & (MW) & (MVAR) \\
        \hline
        1 & 5 & 90 & 30 \\
        2 & 7 & 100 & 35 \\
        3 & 9 & 125 & 50 \\
        \hline
    \end{tabular}
\end{table}

\begin{table}[t!]
    \centering
    \caption{Transmission line parameters for the IEEE 9-bus test system with integrated RES, specifying the resistance ($r_{ij}$), inductance ($x_{ij}$), and susceptance ($b_{ij}$) values of each line connecting the buses based on a 345 kV, 100 MVA base.}
    \label{tab:line_params}
    \begin{tabular}{cccccc}
        \hline
        Sr. & From Bus & To Bus & $r_{ij}$ & $x_{ij}$ & $b_{ij}$ \\
        No. & $i$ & $j$ & (p.u.) & (p.u.) & (p.u.) \\
        \hline
        1 & 1 & 4 & 0 & 0.0576 & 0 \\
        2 & 4 & 5 & 0.017 & 0.092 & 0.158 \\
        3 & 5 & 6 & 0.039 & 0.17 & 0.358 \\
        4 & 3 & 6 & 0 & 0.0586 & 0 \\
        5 & 6 & 7 & 0.0119 & 0.1008 & 0.209 \\
        6 & 7 & 8 & 0.0085 & 0.072 & 0.149 \\
        7 & 8 & 2 & 0 & 0.0625 & 0 \\
        8 & 8 & 9 & 0.032 & 0.161 & 0.306 \\
        9 & 9 & 4 & 0.01 & 0.085 & 0.176 \\
        10 & 7 & 10 & 0.2379 & 1.6189 & 0 \\
        11 & 10 & 11 & 0.01 & 1 & 0 \\
        12 & 7 & 12 & 0.2583 & 1.6552 & 0 \\
        13 & 12 & 13 & 0.01 & 1 & 0 \\
        \hline
    \end{tabular}
\end{table}

\begin{table}[t!]
    \centering
    \caption{Generator cost parameters for the IEEE 9-bus test system with integrated RES, specifying the cost coefficients used in the OPF calculations. The cost function is modeled as a polynomial $C_0 + P_i^GC_1 + {P_i^G}^2C_2$.}
    \label{tab:gencost_params}
    \begin{tabular}{cccc}
        \hline
        Generator & $C_0$ & $C_1$ & $C_2$ \\
         & (euros) & (euros/MW) & (euros/MW$^2$) \\
        \hline
        1 & 150 & 5 & 0.11 \\
        2 & 600 & 1.2 & 0.085 \\
        3 & 335 & 1 & 0.1225 \\
        4 & 8 & 0.3 & 0.0005 \\
        5 & 10 & 0.5 & 0.001 \\
        \hline
    \end{tabular}
\end{table}

\section*{Acknowledgment}
The authors extend their sincere gratitude to J\"ulich Supercomputing Centre for providing computing time on the D-Wave Advantage\texttrademark\, System JUPSI through the J\"ulich UNified Infrastructure for Quantum computing (JUNIQ). The research received support from the Center of Excellence RAISE, which receives funding from the European Union's Horizon 2020–Research and Innovation Framework Programme H2020-INFRAEDI-2019-1 under grant agreement number 951733. The authors also would like to thank Fujitsu Technology Solutions for providing access to the QIIO software\footnote{\url{https://en-portal.research.global.fujitsu.com/kozuchi}} and to Markus Kirsch and Matthieu Parizy for their support and custom extensions of the DADK Python package. Furthermore, the authors acknowledge the support of lab technician Remko Koornneef for his assistance in setting up the RTDS\textsuperscript{\textregistered} laboratory and facilitating hardware communication and integration.

\bibliographystyle{IEEEtran}
\bibliography{ref.bib}

\end{document}